\documentclass[a4paper,fleqn,usenatbib,useAMS]{mnras}
\usepackage{txfonts}
\usepackage{epsfig}
\usepackage{hyperref}
\usepackage{xspace}
\usepackage{lineno}
\usepackage[T1]{fontenc}
\usepackage{ae,aecompl}
\usepackage{multirow}
\usepackage{booktabs}
\usepackage{times}
\usepackage{caption}
\usepackage{subcaption}
\usepackage{{threeparttable}}
\usepackage{multicol}

\usepackage{graphicx}	
\usepackage{amssymb}	
\usepackage{lscape}
\usepackage{placeins}
\usepackage{soul}
\usepackage{longtable}
\usepackage{array, booktabs, longtable, makecell}
\usepackage{rotating}
\usepackage{pdflscape}	

\newcommand{\tes}{\textit{TESS}}

\newcommand{\kep}{\textit{Kepler}}

\newcommand{\jktabs}{\textsc{jktabsdim}}

\newcommand{\spectrum}{\textsc{spectrum}}

\newcommand{\isp}{\textsc{ispec}}

\newcommand{\sigspec}{\textsc{sigspec}}

\newcommand{\kms}{\,km\,s$^{-1}$}

\newcommand{\B}{\textit{B}}
\newcommand{\V}{\textit{V}}

\newcommand{\HH}{\textit{H}}

\newcommand{\spectrumgrid}{\textsc{SPECTRUM\ ATLAS9.Castelli}}

\newcommand{\gaia}{\textit{Gaia}}

\newcommand{\beq}{\begin{equation}}
\newcommand{\eeq}{\end{equation}}

\title[Seismic diagnostics for $\gamma$\,Dor stars]{Exploring the empirical instability strip for $\gamma$ Dor-type stars
	in eclipsing binaries}

\author[ Ö. Çakırlı et al. ]{Ö. Çakırlı$^{1,3}$\thanks{E-mail: \href{mailto:omur.cakirli@ege.edu.tr} {omur.cakirli@ege.edu.tr}}, 
B. Hoyman$^{1,2}$, O. Özdarcan$^{1}$ and S. Bilir$^3$\\
$^1$Ege University, Science Faculty, Astronomy and Space Science Dept., 35100 Bornova, {\.I}zmir, Türkiye \\
$^2$Atatürk Üniversitesi, Fen Fakültesi, Astronomi ve Uzay Bilimler Bölümü, Erzurum, Türkiye \\
$^3$Istanbul University, Faculty of Science, Department of Astronomy and Space Sciences, 34119, Beyazıt, Istanbul, Türkiye
}
\date{Accepted XXX. Received YYY; in original form ZZZ}

\pubyear{2025}

\begin{document}
\label{firstpage}
\pagerange{\pageref{firstpage}--\pageref{lastpage}}
\maketitle

\begin{abstract}
This research paper has light curve modelling, spectroscopy, and detailed asteroseismic studies for 
five detached eclipsing binaries with a $\gamma$\,Dor component that have been found so far by 
the sky surveys. The objective is to study the pulsational characteristics of the oscillating 
stars of the systems, as well as to estimate their absolute parameters and their relation to the 
pulsational frequencies, and enrich the sample size of this kind of system, which has been relatively 
poor up today. The physical properties of these systems are compared with other similar 
cases and the locations of their components are plotted in the Hertzsprung-Russell (HR) diagrams. In order to discover the 
pulsation frequencies, the photometric data are analysed using eclipsing binary modelling methods and 
their residuals are subjected to Fourier analysis. Finally, we evaluated that there is a possible 
relationship between the pulsation periods of the primaries and the orbital period ($P_{\rm orb}$), the 
force exerted per gram on the surface of the pulsating star ($f$/M$_{\rm puls}$), and the fractional radius 
of the pulsating star ($r_{\rm puls}$) of the systems, using a sample of 39 eclipsing binary systems with 
$\gamma$ Dor type primaries. We utilised them to construct a unique observational $\gamma$\,Dor instability 
strip, delineated by a lower limit of 6\,850\ K at the red edge and an upper limit of 7\,360\ K at the blue 
edge on the zero-age main sequence. The majority of 39 pure $\gamma$\,Dor stars are located within the 
region that is covered by this observational strip.
\end{abstract}


\section{Introduction}\label{intro}
The study of stellar oscillations is a powerful technique for probing the interior structures of stars using 
periodic stellar oscillations \citep{2010aste.book.....A,2021RvMP...93a5001A}. Oscillations can be observed 
through photometric methods  \citep{2013pss4.book..207H}, and one can uncover the pulsation frequencies by analysing 
the amplitude spectrum. From that point, you can ascertain various parameters including mass, age, rotation, and 
distance \citep[e.g.,][]{2014MNRAS.444..102K,2018MNRAS.474.2774S}. So, we focus on $\gamma$\,Doradus stars, which 
are A- to F-type main-sequence stars with masses that are usually between 1.4 and 2.0 $M_{\odot}$ 
\citep{1999PASP..111..840K}. Pulsations in $\gamma$\,Dor stars are caused by gravity modes with high radial order 
and low degree. A typical pulsation period is between 0.3 
and 3\,days \citep[e.g.][]{1994MNRAS.270..905B,1999PASP..111..840K,2018MNRAS.477.2183S,
2018A&A...618A..24V,2019MNRAS.487..782L}. In the vicinity of the core, gravity modes have the maximum mode 
energy \citep{2015ApJ...810...16T,2016A&A...593A.120V}. Because of this, $\gamma$\,Dor stars let us look 
into the stellar interiors. However, there is still some controversy on the excitation process of $\gamma$\,Dor 
stars. Both \citet{2000ApJ...542L..57G} and \citet{2005A&A...435..927D} indicated that the convective flux 
blocking mechanism that acts at the base of the envelope convection zone is responsible for the excitation 
of the g-mode pulsations in the envelope. \citet{2016MNRAS.457.3163X} revealed that the radiative $\kappa$ 
mechanism plays an important role in $\gamma$\,Dor stars, whereas the connection between convection 
and oscillations dominates in cool stars. The instability strip of $\gamma$\,Dor stars is between the sun-like 
stars and the $\delta$\,Scuti stars. It overlaps with the red edge of the instability strip of $\delta$\,Scuti \citep{2005A&A...435..927D,2009AIPC.1170..477B,2013MNRAS.429.2500B}. It was difficult to identify the pulsations 
of $\gamma$\,Dor stars using ground-based observations because of the daily aliasing and the modest amplitudes 
of the oscillations.  

Binary systems underpin numerous key relationships utilised in asteroseismology. Masses and radii are rarely 
determined with more precision than through the analysis of orbital dynamics and the depths of eclipses. Therefore, 
$\gamma$ Doradus-type pulsating stars in binary systems present significant opportunities for research. They 
provide a dual opportunity to enhance our knowledge of the interior physics of stars through the utilisation 
of gravity modes. Pulsations are crucial for helping probe deep regions within a star, facilitating the 
comprehension of stellar rotation, convective core boundary mixing, envelope mixing, opacity, and magnetic 
fields. To achieve this objective, we have conducted a survey aimed at increasing the number of $\gamma$ Doradus stars 
found in eclipsing binaries. This work is a continuation of the comprehensive investigation of detached eclipsing 
binary stars with a $\gamma$\,Dor component \citep[see][]{2020MNRAS.491.5980H}. We report analyses of five detached eclipsing 
binaries. All of these systems have previously been described by \citet{2020A&A...643A.162S,2019A&A...630A.106G,
2019AJ....158..106C} as "pulsating" eclipsing binaries, but there has been no detailed study of any of them nor 
any classification of their spectroscopy and Roche geometry by models to date. This lack of information is due 
to the fact that these systems have not yet been studied in depth. So, this work adds a lot to the list of 
detached eclipsing binaries with $\gamma$\,Dor components whose absolute stellar properties are well-determined. For 
the selected systems, space-based photometric and high-resolution spectroscopic data have been collected 
(see Section\,\ref{data}) in order to: a) perform detailed light and atmospheric modelling (Section\,\ref{ravespan}, 
Section\,\ref{spec_analysis}), b) calculate the stellar parameters with accuracy 
(Section\,\ref{pywd}), c) determining the evolutionary status to compare with theoretical models (Section\,\ref{sec_iso}), 
and d) detect the most powerful pulsation frequencies and the influence of absolute parameters on the evolution 
of the system (Section\,\ref{puls}). The properties of the systems as well as the pulsation properties of their 
oscillating stars are compared with others of similar type. Finally, conclusions are given in Section\,\ref{conc}. 
%

%
%
%
\begin{table*}
	\centering
	\caption{Literature information about the targets from this work.}
	\label{tab:cato_dat}
	\begin{tabular}{lccccc}
	\hline
\hline
Catalog IDs & $\alpha_{2000}$& $\delta_{2000}$ & $P^a$~(d) & $T_0$$^a$ (BJD-2450000) & $K_p$ / \tes\ (mag)\\
\hline
KIC\,4932691	& 19:36:57.63 & +40:00:26.84 & 18.11208(7)	& 54967.7127 & 13.627$^b$ \\
TIC\,7695666	& 04:22:35.22 & -41:29:00.13 & 17.55370(93) & 58429.2434 & 10.127$^c$ \\
TIC\,410418820	& 05:30:13.88 & -84:47:06.37 & 8.569526(49) & 58633.4085 &  6.213$^c$ \\	
EPIC\,203929178	& 16:49:41.47 & -24:22:25.64 & 2.307865(9)$^d$ & 56894.7463 & 11.160$^b$ \\
TIC\,213047427	& 05:01:24.93 & -05:41:43.52 & 8.800106(21) & 58455.3916 &  8.325$^c$ \\
\hline
	\end{tabular}\\
$^{a}$ Orbital period of the eclipsing binary, where $T_0$ is the mid-time of the primary eclipse 
(\url{http://keplerebs.villanova.edu/} and \tes\; \citealt{2022ApJS..258...16P}); $^b$ Estimated 
magnitude in \kep\ band; $^c$\tes\ magnitude \citep{2022ApJS..258...16P}; $^d$ The orbital period of 
the EPIC\,203929178 is accepted as double the value stated in the literature.
\end{table*}
%
%
%

\section{Data and Methods}\label{data}
\subsection{Target selection}
The detection of pure $\gamma$ Dor-type stars—those that clearly match the criteria of $\gamma$ Dor based on their 
position on the HR diagram and their pulsation traits—is quite rare. Conversely, a notable quantity of stars 
exhibit hybrid ($\delta$ Scuti + $\gamma$ Dor) type changes. These changes are 
frequently observed in stars that are components of a binary star system or in single stars. However, 
our research aims to concentrate on the star components of a binary system that demonstrates 
$\gamma$ Dor-type pulsation modes. Our investigation focused on a specific set of 
systems, excluding hybrid systems of the $\delta$ Scuti + $\gamma$ Dor type, with the 
third body effect, and rare stars like white dwarfs or other exotic and evolved stars.

The examination of these systems through the use of spectroscopy has been limited up to the present 
study, with the exception of the analysis that was carried out by photometry. The objects chosen for 
analysis were based on their previously detected spectra and space based photometric data, without any 
additional selection criteria being applied. We aimed to provide an accurate representation of their 
orbital cycles and determine their orbital parameters with a margin of error of around 5\%. Two 
different types of data were utilised in this investigation for its purposes. The photometric light 
curves sourced from the Transiting Exoplanet Survey Satellite (\tes; \citealt{2015JATIS...1a4003R}) 
and \kep\ \citep{2010Sci...327..977B} public data. Following closely behind are the spectral data that 
we employ in order to ascertain the atmospheric characteristics and radial velocities of the stars that 
we have considered. Data-related information will be presented in more detail in subsequent sections of 
this paper. For now, Table~\ref{tab:cato_dat} presents a summary of the sample stars, 
including their observational characteristics in detail.
%

%
%
\begin{table*}
\caption{Summary of \tes\ and \kep\ observations.}
\begin{tabular}{lclc}
\hline
 \hline
\kep\ / \tes\ IDs		&Sectors&Observing Start/End (UT)& Exposure time/type\\
\hline
KIC\,4932691			&---					    &2009 May 13/2013 May 11 &1800/LC  \\
TIC\,7695666			&4,5					    &2018 Oct 19/2018 Dec 11 &120/SPOC \\
TIC\,410418820			&12,13,27,38,39,65,66,67    &2019 May 21/2023 Jul 29 &120/SPOC \\
EPIC\,203929178			&---					    &2014 Aug 24/2014 Nov 10 &1800/ K2 \\
TIC\,213047427			&5,32					    &2018 Nov 15/2020 Dec 16 &120/SPOC \\
\hline
\end{tabular}\\
\begin{tablenotes}
\footnotesize
\item
Based on the data available on MAST\footnote{\url{https://archive.stsci.edu/}} as of 
November 1, 2023. The TIC numbers are used to organise the systems in this list. The 
\tes\ observing baseline includes the start and end dates, which are listed in the 
"Observing start/end" column. However, it is important to note that the systems may 
not have been observed continuously throughout this entire period. The \tes~light 
curves for all systems are observed with a 120-second exposure time cadence. 
For the KIC\,4932691 and EPIC\,203929178, we use \kep\ and K2 data, respectively.

	\end{tablenotes}
\label{tab:tess_summary}
\end{table*} 
%
%

\subsection{Photometry}\label{photometry}
We retrieved the \tes\ light curves from the Science Processing Operations Centre pipeline 
\citep[SPOC;][]{2016SPIE.9913E..3EJ}, which were downloaded using the \textsc{lightkurve} module 
under \textsc{python} environment \citep{2018ascl.soft12013L} from the Mikulski Archive for 
Space Telescopes (MAST\footnote{\url{https:/ mast.stsci.edu/portal/Mashup/Clients/Mast/Portal.html}}). We 
conducted a visual examination of the light curves for each object and chose data sources characterised 
by minimal scatter and clear total eclipsing signals. For the analysis, we utilised the \textsc{PDCSAP} 
(Pre-search Data Conditioning Simple Aperture Photometry) flux, excluding only those data points that 
had a non-zero quality flag. When multiple sectors were available in the MAST archive for a target, we 
merged all the individual light curves to create a unified and continuous multi-sector light curve. The 
stitch method in the \textsc{lightkurve} package is employed to concatenate and normalise all the 
sectors. To investigate potential blends and contaminating sources in the apertures, we utilised 
again \textsc{lightkurve}  to plot the Target Pixel Files (TPF) for each central star in the sample. The 
examination of the TPFs holds particular significance in crowded fields, as the comparatively large pixel 
size of the \tes\ CCDs may result in considerable photometric contamination from nearest sources. Overall, we 
identified our central targets that exhibited no significant contamination from adjacent sources within 
the aperture mask. We are confident that the level of contamination, whether negligible or minimal, has 
not influenced our light curve analysis.

We also utilised the photometric light curve from space surveys, such as the \kep\ mission and the K2 
mission of the Kepler Space Telescope, to analyse the light curve of the two systems, since \tes\ did not 
observe EPIC\,203929178 and \tes\ data of KIC\,4932691 is not as good as \kep\ data for pulsation analysis. The 
name K2 may have been selected because it represents the second \kep\ mission or because the Kepler spacecraft 
is working with only two remaining reaction wheels \citep{2014PASP..126..398H}. The primary objective of the K2 is 
to observe near the ecliptic plane, capturing images of various target fields. We have conducted our analysis 
utilising all available \kep\ and $EPIC$/K2 data from the MAST Archive. We obtained $tar$ files that include 
all the long-cadence (LC) light-curve files for KIC\,4932691, and we also acquired the complete K2 input 
catalogue (EPIC) for Campaign-2 of EPIC\,203929178.

\subsection{Spectroscopy}
The optical spectra consist of data acquired simultaneously with the survey observations 
as part of the ESO follow-up programmes (ESO Science Archive 
Facility\footnote{\url{http://archive.eso.org/wdb/wdb/adp/phase3_main/form}}), along with 
additional spectra taken at different moments. The FEROS \citep[The Fiber-fed Extended Range Optical Spectrograph, $R\sim$ 48\,000;][]{1999Msngr..95....8K} 
echelle spectra, obtained from the 0100.D 0339(B) follow-up programme, have a phase distribution that 
is adequately represented. The UVES \citep[Ultraviolet and Visual Echelle Spectrograph, R$\sim$51\, 000;][]{2000SPIE.4008..534D} 
spectra, obtained from the 01094.D 0190(A) follow-up programme. The observations were made 
using the Dichroic 2 instrument mode, which provided spectral ranges of 373-499 nm (blue arm), 
567-757 nm (red arm lower), and 767-945 nm (red arm upper). We also retrieved HARPS 
\citep[High Accuracy Radial velocity Planet Searcher;][]{2003Msngr.114...20M}  archival data for our targets listed in Table~\ref{tab:target_RVs}. The 
preference for ESO\,3.6–metre telescope (R$\sim$80\,000, in the high efficiency mode; EGGS and $R \sim$115\,000, in 
the high RV accuracy mode; HAM) is motivated by its characteristics: large wavelength range (the optical 
spectral region from $\sim$3\,800 to $\sim$6\,900\ \AA\ in only one exposure), high resolution, and high spectral 
stability, which makes it suitable for detecting narrow absorption features in a wide variety of spectral lines.

The optical spectra are supplemented by a series of near-infrared APOGEE (The Apache Point Observatory Galactic Evolution Experiment) high-resolution echelle 
spectra. APOGEE is a survey of \textit{Milky\,Way} stars using a multi-object, fiber-fed, near infared (NIR) 
spectrograph that records most of the \HH-band with resolving power $R\sim$22\,500 on three non-overlapping 
detectors (blue detector: 15\,145-15\,810 \AA, green detector: 15\,860-16\,430 \AA, red detector: 
16\,480-16\,950 \AA). The APOGEE instruments operate on the Sloan 2.5m telescope \citep{2006AJ....131.2332G} 
at Apache Point Observatory (APO, APOGEE-N). \cite{2017AJ....154...94M}, provided a detailed overview of the 
APOGEE survey, and \cite{2015AJ....150..173N} described the data reduction process. The information 
that is given in Table~\ref{tab:target_RVs} pertains to the total number, the instrument that was used to 
make the observations, and the signal-to-noise ratio (SNR) of the spectra.

The optical spectra were normalised using the main functionalities for spectra treatment integrated 
into Python-based code \isp\ (the Integrated SPECtroscopic framework; \citealt{2014A&A...569A.111B}), which 
cover the fundamental aspects. This process is known as \textit{continuum\,normalisation}. The technique 
involves finding the continuum points of a spectrum by applying a median and maximum filter with varying 
window sizes. The former method reduces the impact of noise, while the later method disregards the more 
significant changes associated with absorption lines. As a result, the continuum will be slightly shifted 
upwards or downwards based on the values of these parameters. Subsequently, the user can select either a 
polynomial or a set of splines to represent the continuum. Ultimately, the spectrum is normalised by 
dividing all the fluxes by the selected model. After normalising the spectra, we proceed to eliminate 
any erroneous spectral features caused by tellurics through a process known as \textit{de-spiking}. This 
step is crucial in ensuring the accuracy of our data.

\section{Analysis}\label{analysis}
\subsection{Radial velocity measurements and orbital solutions}\label{ravespan}
The radial velocity of the host star undergoes temporal variations in a binary system due to 
the presence of a companion star. Through careful analysis of the radial velocity data, valuable 
insights can be gained regarding the relative masses of the host and companion, as well as 
important orbital parameters such as period and eccentricity. In this section, we will explain 
the method of establishing the inference of radial velocity measurements within a Markov Chain 
Monte Carlo (MCMC; \citealt{2010CAMCS...5...65G}) framework. This method has been widely used 
in astrophysics for searching, optimising, and sampling probability distributions. We used the 
normalised spectra and adopted the \spectrum\ radiative transfer code 
\citep{1994AJ....107..742G}, and the stellar atmospheric model ATLAS9.Castelli model 
atmospheres \citep{2003IAUS..210P.A20C}, and the line lists provided alongside the \spectrum\ 
code. These were modified to be compatible with \isp\ software. We specified a range of parameters 
for the synthesised spectra, including $T \rm_{eff}$, $\log g$, [M/H], $v\sin i_{\rm obs}$, $V_r$, 
and $lf$ (light fraction). Here, $T\rm_{eff}$, $\log g$, [M/H] denote the atmosphere properties of the star, namely 
the effective temperature, surface gravity, and metal abundance, respectively. Both the projected rotation 
and the radial velocity of the star are denoted by the symbols  $v\sin i_{\rm obs}$ and $V_{\rm r}$ 
respectively. Lastly, the $lf$ value symbolises the light fraction obtained by introducing 
Gaussian noise to the entire synthetic spectrum.

%
%
\begin{figure}
	\includegraphics[width=0.45\textwidth]{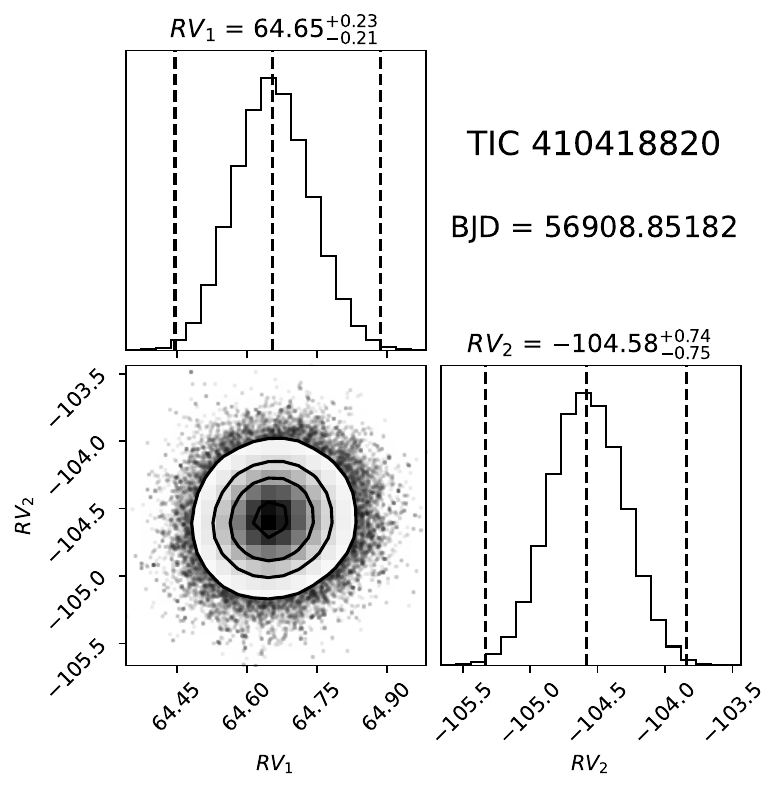}
	\caption{Illustrative corner plots displaying the observed radial velocities for the spectra of TIC\,410418820.}
	\label{mcmc_rv}
\end{figure}
%
%

%
%
\begin{figure*}
	\includegraphics[width=0.90\textwidth]{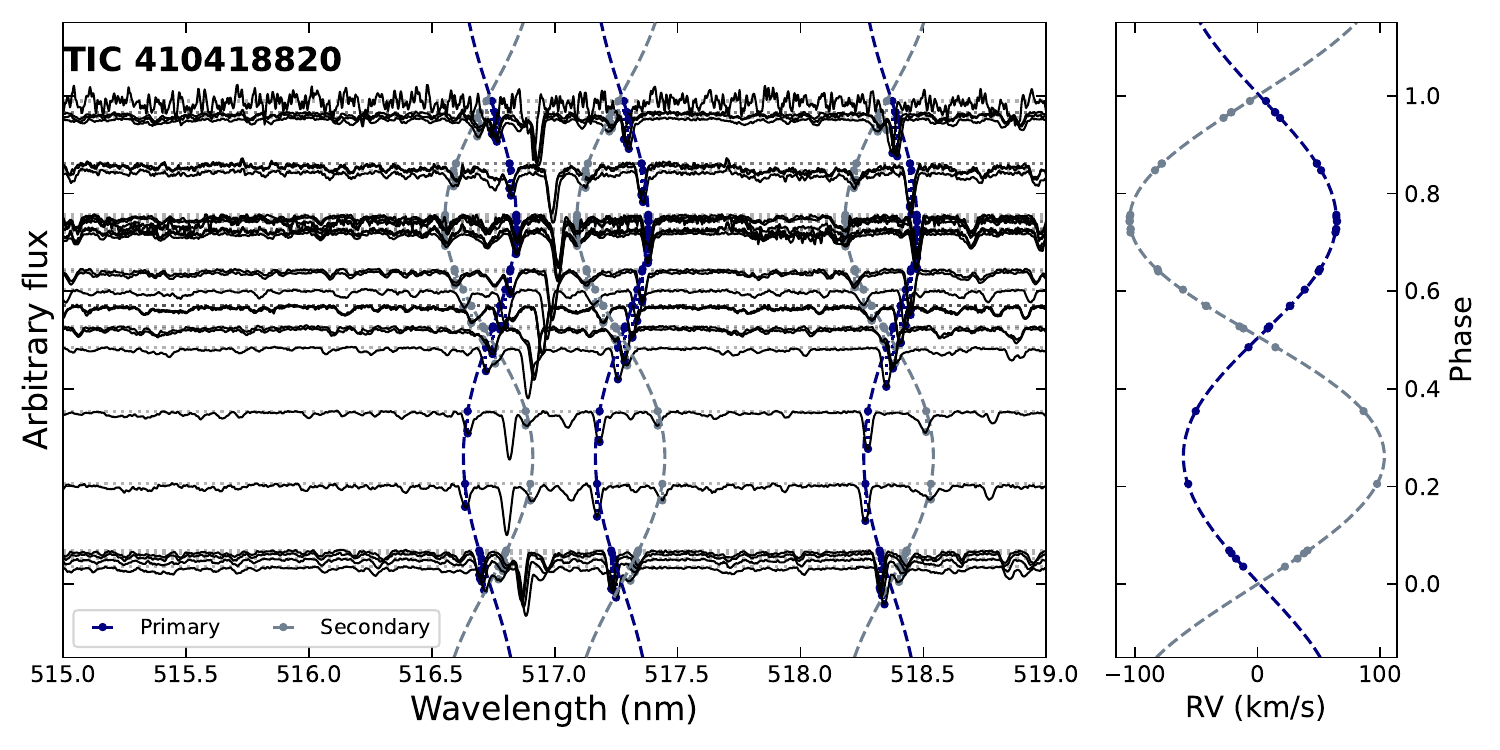}
	\caption{The left panel shows the development of absorption lines with a second 
		blue component that follows a cycle of orbital period for TIC\,410418820. The 
		variation of the lines is indicated by blue and grey dashed lines. The right 
		panel displays the radial velocity of the absorption lines. The blue filled 
		circles on the radial velocity plot indicate the primary, while the grey filled 
		circles represent the secondary.
	}
	\label{phased_mcmc_rv}
\end{figure*}
%
%

We obtained the normalised synthetic spectra of single stars using \isp\, based on these parameters. We 
merged the synthetic spectra of two individual stars without introducing any additional noise 
to generate binary spectra. Upon creating the synthetic spectra for each individual component, 
we started the process of measuring the radial velocity. In order to do this, we identified the 
spectral areas that were not influenced by the telluric lines, and we found that both of the 
components were able to be recognised with relative ease. Through the use of these areas, we 
were able to crop the spectra that previously were seen. The synthetic composite spectra were 
created by adjusting the components' synthetic spectra based on their expected radial velocities. 
The light contributions obtained from preliminary binary modelling were used to scale the spectra. The 
observed and theoretical spectra were then compared to determine how well they corresponded. The 
MCMC algorithm allowed for flexibility in adjusting the radial velocity of the components in 
order to find the optimal fitting solution. The algorithm assumes that the likelihood function 
for this process is represented by $ln(p)$=$\chi^2$/2. The MCMC solutions are executed using 24 
walkers, with each walker taking 5\,000 steps for every observed spectrum.

Figure~\ref{mcmc_rv} displays an exemplary corner plot showcasing the system's MCMC solutions. 
The measured radial velocities are determined by analysing the sample data of each solution and 
calculating the lower error, median value, and upper error based on the 3\%, 50\%, and 97\% 
quantiles. All of the radial velocities that were measured for each of the systems are presented 
in Table~\ref{tab:RVs}.

It is evident from the spectroscopic variations that the absorption lines undergo periodic changes, 
aligning with the orbital period of the system. An exemplary demonstration of this principle may 
be seen in the spectrum observations of TIC\,410418820, for example. Figure~\ref{phased_mcmc_rv} 
clearly demonstrates the spectroscopic variation through the multitude of absorption lines. The 
periodic shift of the isolated absorption lines profiles across numerous consecutive runs is 
also illustrated in Figure~\ref{phased_mcmc_rv}, where the absorption lines from the systems are 
clearly distinguished and displaced along the orbital phases. Furthermore, the right panel of 
Figure~\ref{phased_mcmc_rv} displays the measured radial velocities with the corresponding radial 
velocity model that has been fitted. It seems that all the stars that have undergone spectrum 
analysis have completed a full cycle, to different extents. However, the reliability of the 
analysis is not affected in any way by the fact that there are intervals between each step.

\begin{figure}
\includegraphics[width=0.45\textwidth]{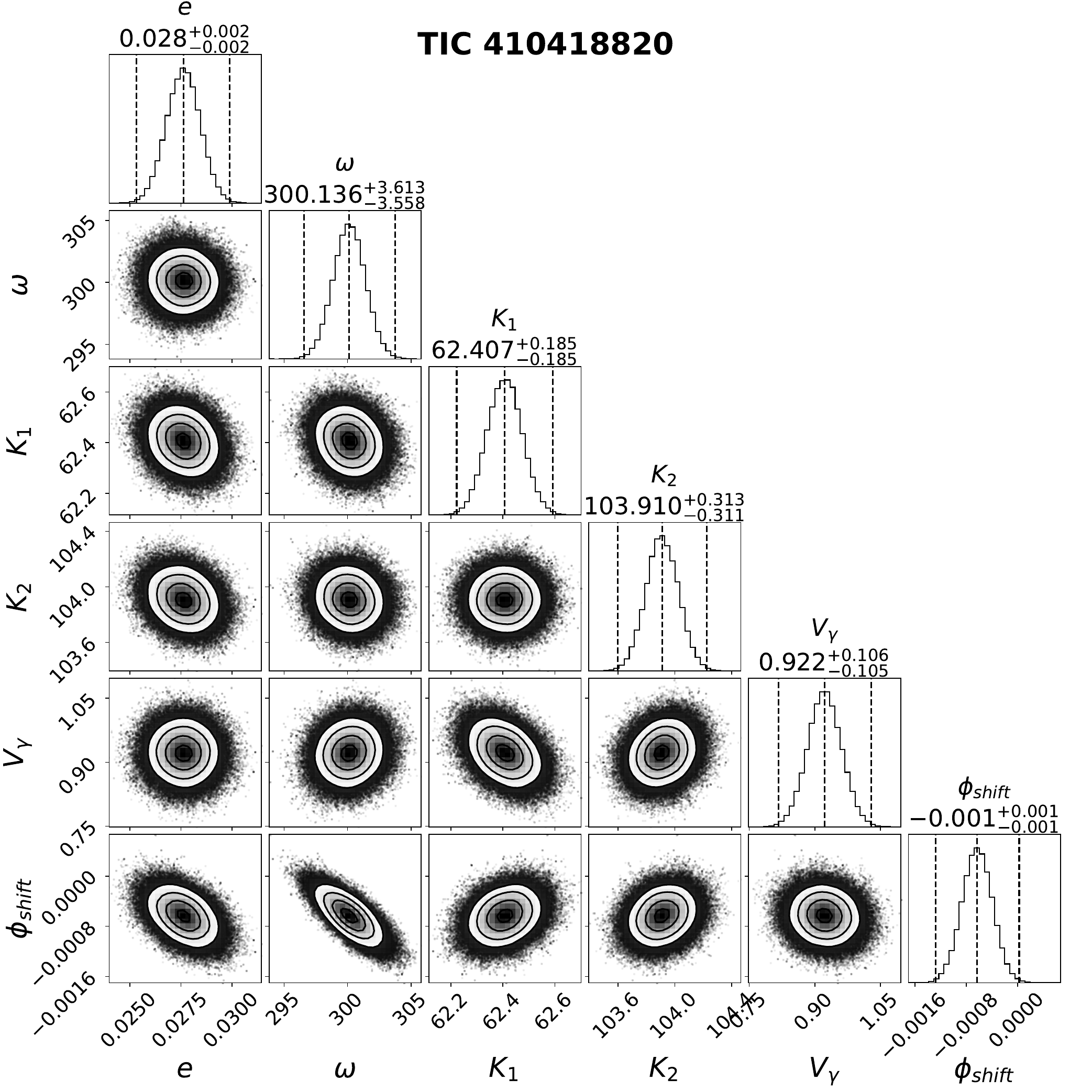}
	\caption{Corner plot showing the distributions and correlations of the TIC\,410418820 orbital parameters 
from the ﬁt of the radial velocity data. Contour levels correspond to 1, 2, and 3$\sigma$, and the histograms 
on the diagonal represent the posterior distribution for each parameter, with the mode and internal 99\% confidence 
levels indicated. More realistic errors are discussed in the text.
	}
	\label{mcmc_rob}
\end{figure}

\begin{figure*}
\center
\includegraphics[width=5.6 cm]{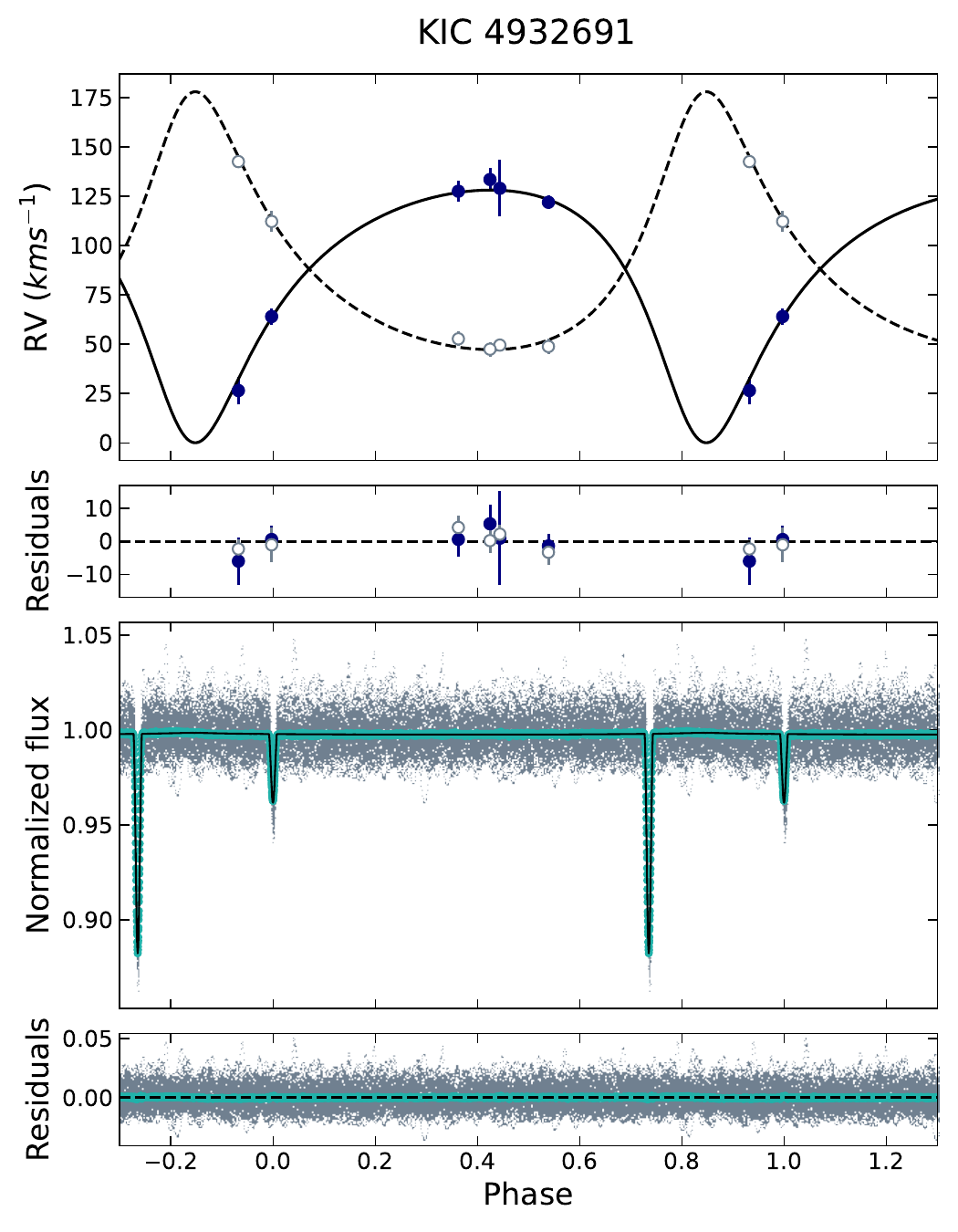}
\includegraphics[width=5.6 cm]{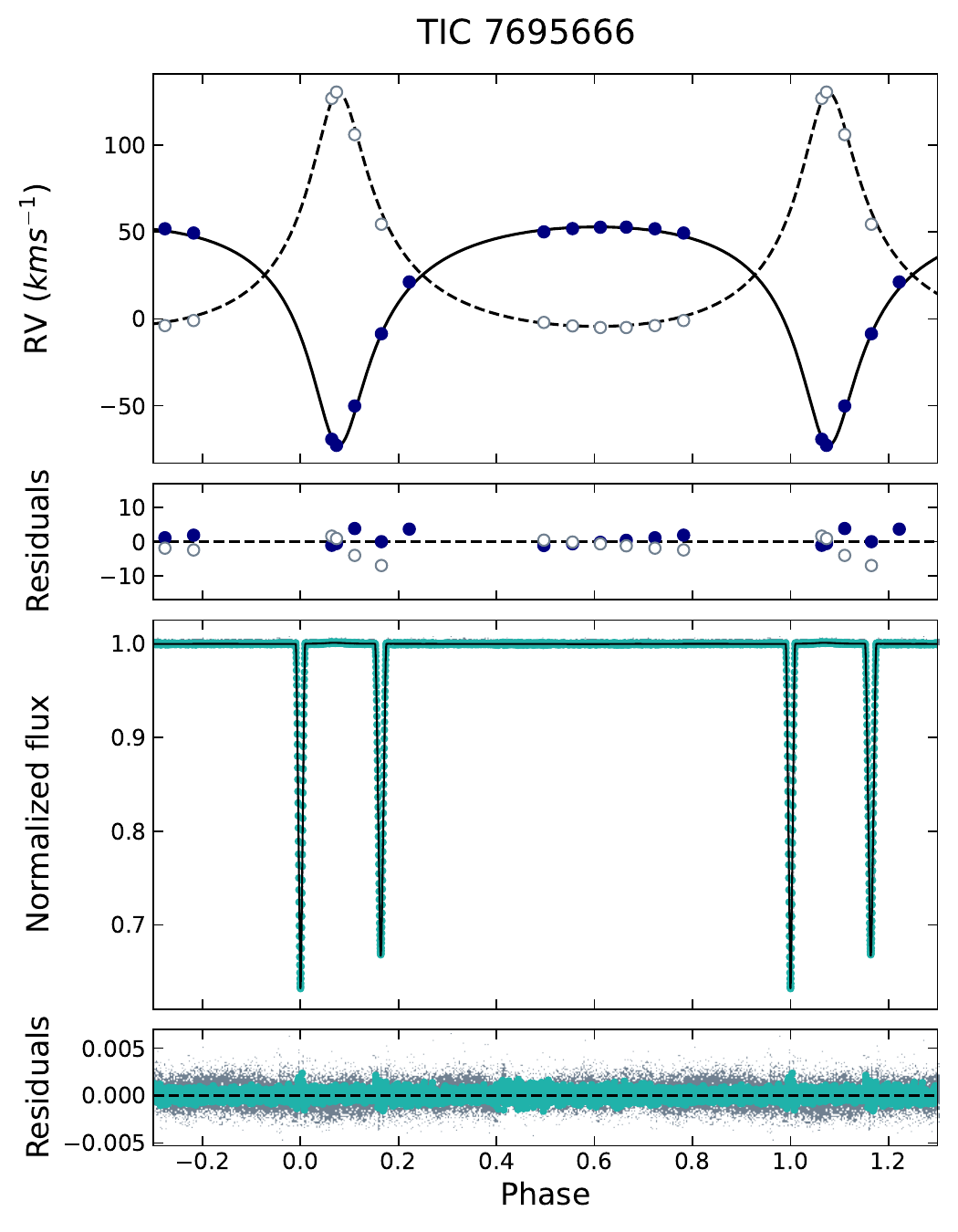}
\includegraphics[width=5.6 cm]{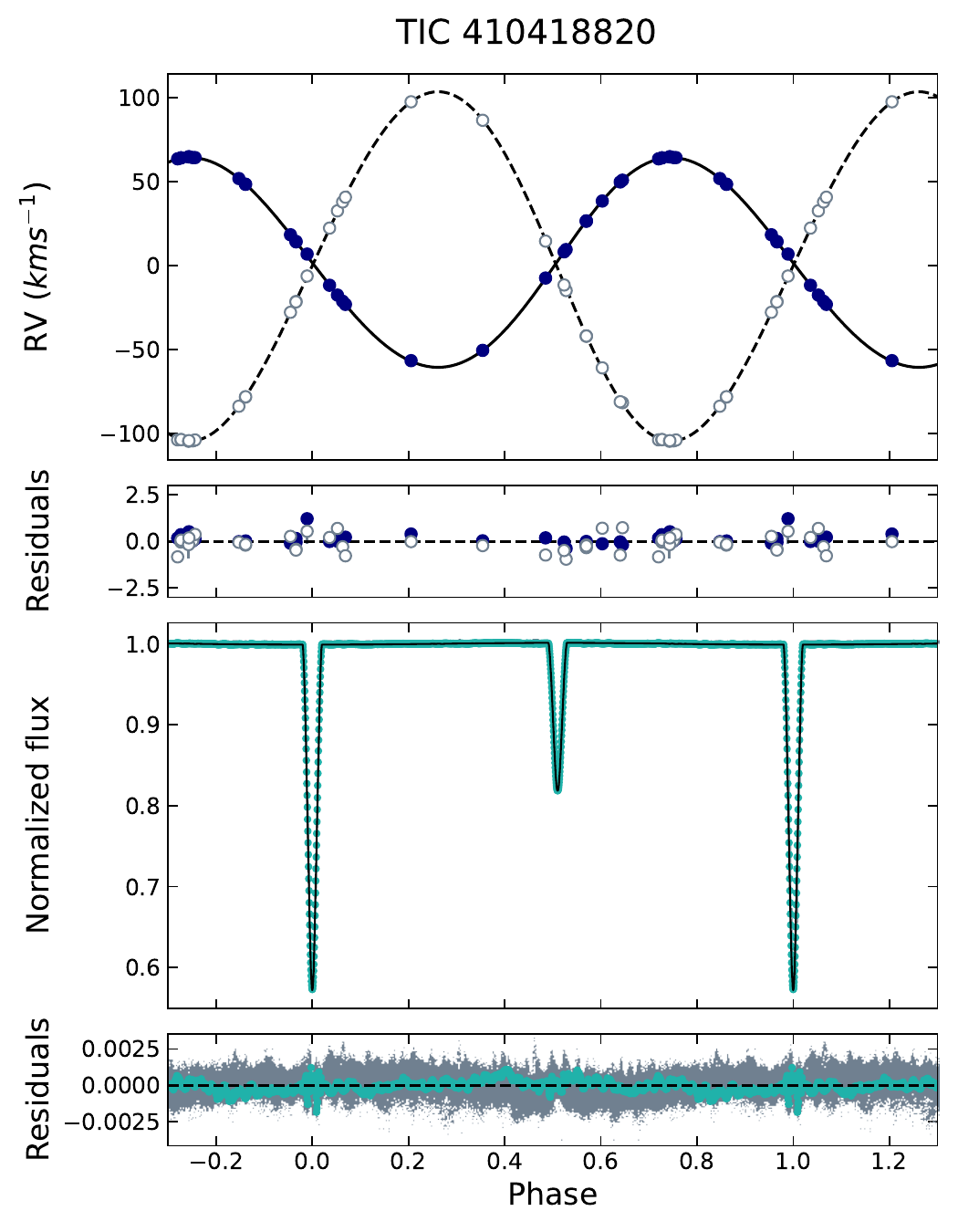}
\includegraphics[width=5.6 cm]{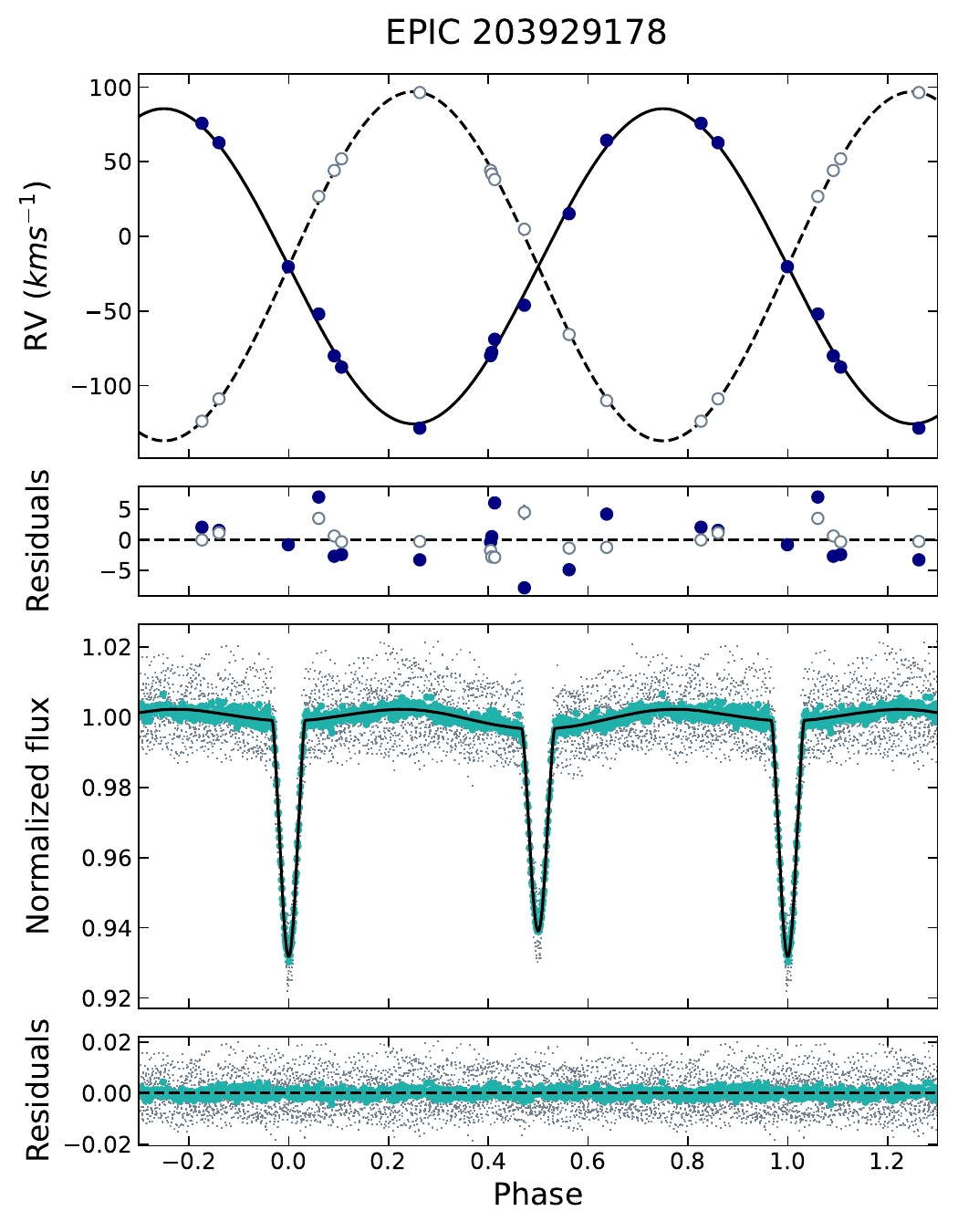}
\includegraphics[width=5.6 cm]{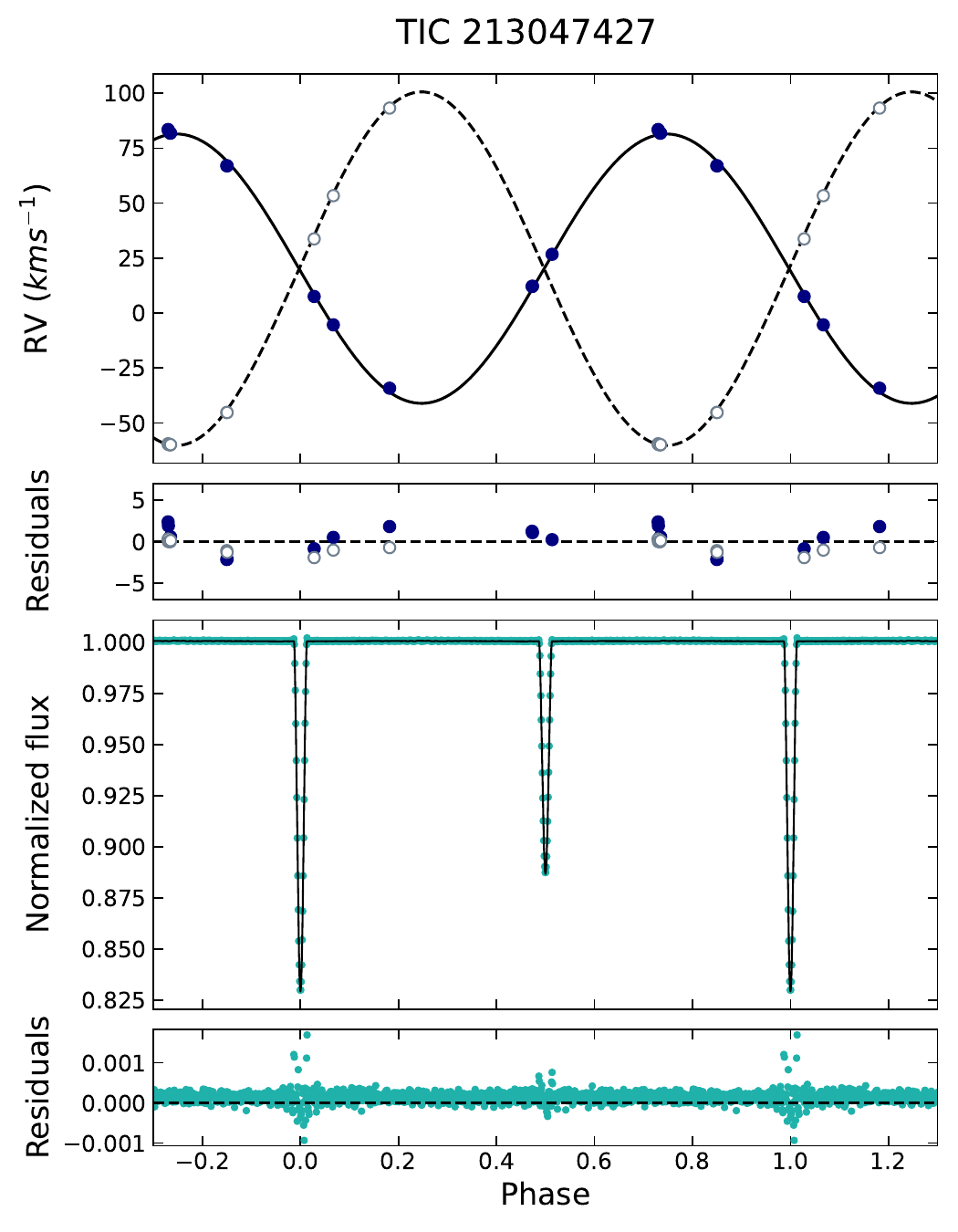}
\caption{The joint analysis successfully fits the \tes~photometry and radial velocity time series 
of the target stars. For each target system, the upper two panels display the radial velocity curves and residuals from the model, while the lower two panels display the observational light curve, with binned mean curves overplotted in different colour and residuals. The models that exhibit the best fit are represented 
by continuous lines in black. The light curve model has been fitted to the entirety of the \tes~photometry 
dataset. The term '1.0' refers to the mid-time of the deeper eclipse phase. The filled symbols on the 
radial velocity plot represent the primary, while the open symbols represent the secondary, as shown 
in the upper panel. The best-fitting models for the primary and secondary stars are represented by 
continuous and dotted lines, respectively, in black colour. }
\label{LC_RV_plot}
\end{figure*}

By measuring the radial velocities of the components, we are able to determine the orbital 
solutions of each system. This is done using the MCMC algorithm, which efficiently explores 
parameter space and provides probability distributions of orbital parameters. We utilise the 
\textsc{emcee} code, which is a Python version of the Goodman-Weare affine invariant 
\citep{2010CAMCS...5...65G} MCMC ensemble sampler developed by \citet{2018ApJS..236...11H}. The 
data, the prior distributions of the model parameters, the likelihood function, and the beginning 
locations of the random walkers are the key elements that are used to enter the code into the 
computer. Before delivering the posterior distributions of model parameters as outputs, the 
algorithm first determines the likelihood of various solutions, then executes jumps and explores 
the parameter space for a certain number of steps, and then returns the results. The solution 
was performed with the assumption that the likelihood function was $\rm ln(p)$ and several parameters 
including orbital eccentricity ($e$), argument of periastron ($\omega$), radial velocity amplitude 
of the components ($K_{1,2}$), radial velocity of the centre of mass ($V_{\rm \gamma}$), and arbitrary 
phase shift ($\phi_{\rm shift}$) were considered as adjustable. The orbital period ($P$) and mid-time of the 
primary minimum ($T_0$) were obtained from light curve solutions and remained constant 
throughout. The MCMC solutions are executed with 48 walkers, each taking 10\,000 steps per 
walker for every system. The corner plots for each solution can be found in 
Figure~\ref{mcmc_rob}, while the fitted parameters and other derived quantities for the 
best-fit model are provided in Table~\ref{tab:salt_table}. Graphical representations of observations and models
are shown in Figure~\ref{LC_RV_plot}. The parameter determination and error estimation 
method for the radial velocities were applied in the same manner. The sample sets were 
calculated using the final MCMC samples to determine the parameters and errors.
%

%
%
%
%

%
%
\begin{figure*}
	\center
	\includegraphics[width=\textwidth]{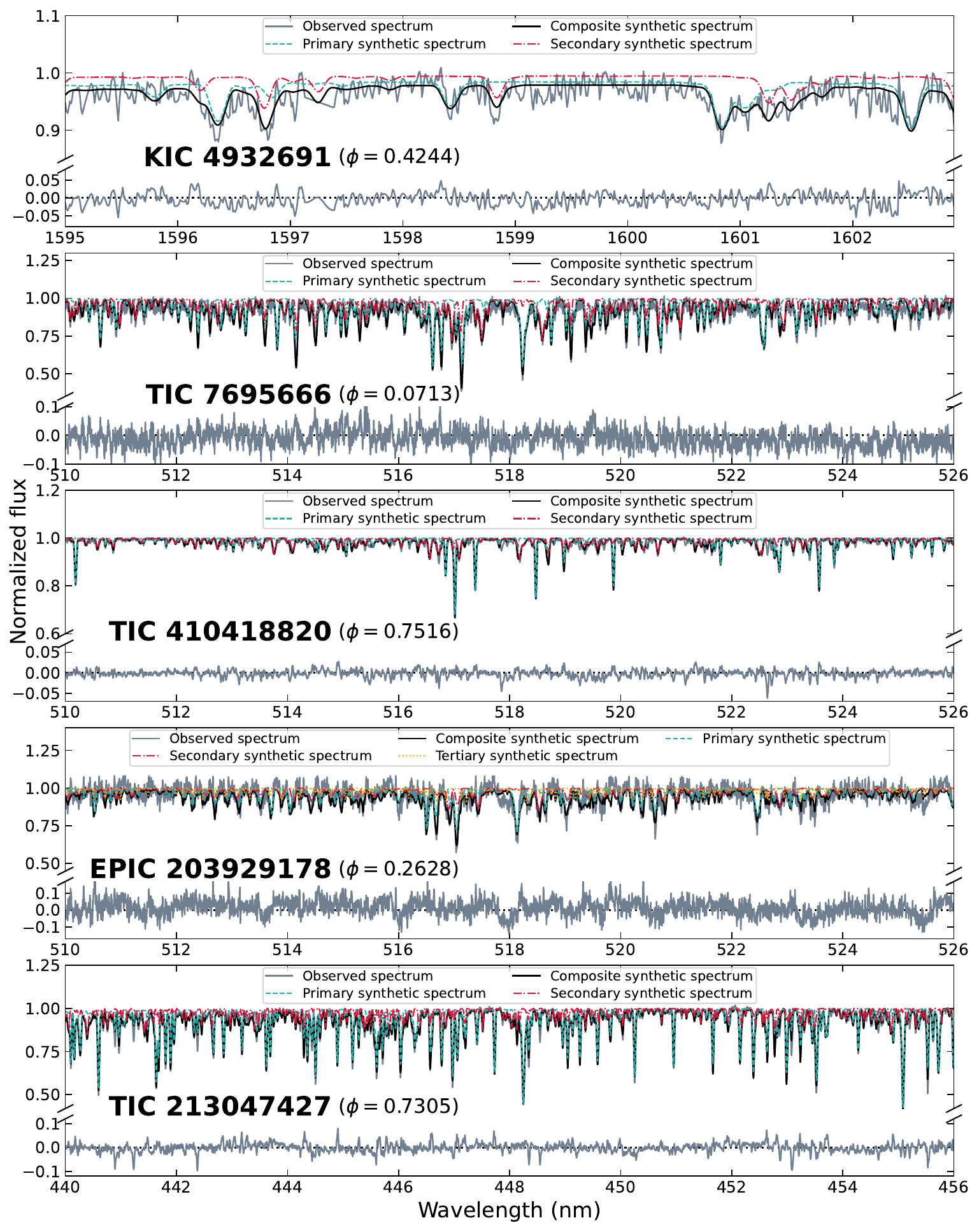}
\caption{The composite spectra of the systems are compared to a combined synthetic model, which are represented 
by solid lines for the observed and model composite spectra, dashed, dotted and dash-dotted lines for individual component spectra.
}
	\label{mcmc_tayf}
\end{figure*}
%
%

\subsection{Atmospheric parameters}\label{spec_analysis}
The high-resolution optical spectra contain crucial indications of stellar temperature 
and other physical parameters. An iterative approach is required to thoroughly examine 
the spectral data and obtain accurate atmospheric parameters when analysing binary systems 
using the model spectra. This thorough examination is crucial because of the importance 
of the data at hand. We aim to achieve the optimal overall fit of the observed data, taking 
into account the sensitivity of the diagnostics to the stellar parameters. The process 
begins by estimating the stellar parameters using the spectral type of the target. A 
preliminary model of the stellar atmosphere is then generated, and its resulting spectrum 
is compared to the observed data. In order to determine fundamental atmospheric characteristics 
such as $T_{\rm eff}$, $\log g$, $v\sin i_{\rm obs}$ and [M/H], it is crucial to obtain a synthetic 
spectrum that closely resembles the observed spectrum. But the observed spectra pose a 
challenge to the fitting procedure due to their combination of two stellar spectra. The 
synthetic composite spectrum is achieved by calculating a synthetic spectrum for each 
component, evaluating their impact on the overall luminosity, adjusting the component 
spectra accordingly shifted depend on radial velocities, and then comparing them with 
the observed spectra to identify the optimal combination of parameters. These comparisons 
are conducted within the global uncertainty range, without any automated minimising 
methods. The model parameters are adjusted based on the outcome of the initial comparison, 
resulting in the calculation of a new atmosphere model. The procedure is repeated until 
the observations fit well with the normalised line spectrum. Figure~\ref{mcmc_tayf} 
displays the ultimate alignment of the normalised line spectrum of the systems as 
an illustration. In order to tackle this difficulty, it is necessary for us to focus 
on enhancing our code.

The code is written in Python programming language and utilises \isp\ subroutines 
\citep{2014A&A...569A.111B, 2019MNRAS.486.2075B} and pre-generated synthetic spectrum 
libraries to generate a synthetic spectrum for each component. The spectra undergo a 
shift according to the radial velocity of each component and are subsequently scaled 
to account for their respective contributions to the overall luminosity. The end result 
is a "composite spectrum". The synthetic composite spectrum can be adjusted to cover 
the entire observed spectrum or focused on specific segments to enhance computational 
efficiency and optimise results. Afterwards, we analyse the generated spectra by 
comparing them to the observed spectra in order to determine the correlation between the 
two. The code employs the \textsc{emcee} package, which is a Python implementation of 
affine-invariant ensemble samplers for MCMC \citep{2010CAMCS...5...65G}. This procedure 
is executed multiple times to improve the independent atmospheric parameters and obtain 
a set of parameters that result in the most accurate composite spectrum. The parameters 
used in this process encompass a range of variables, such as 
$T_{\rm eff\,1,2}$, $\log g_{1,2}$, $\rm [M/H]_{1,2}$, $v \sin i_{1,2}$, alpha element 
abundance ($\alpha_{1,2}$), micro and macroturbulence velocities ($v_{\rm mic,1{},2}$, $v_{\rm  mac,1{},2}$), 
limb darkening coefficients ($\rm LD_{1,2}$), spectral resolutions ($R_{1,2}$), radial 
velocities ($RV_{1,2}$), and the fractional contribution of the components to the total 
luminosity ($\rm LFrac_{1,2}$). The solution process involves keeping certain parameters 
constant, while others are subject to defined relationships that impose constraints on the 
solution. Furthermore, it is possible to ascertain the variance ranges for independent 
parameters, and the solution has been narrowed down to this particular range.

%
%
\begin{figure*}
\includegraphics[width=17.5 cm]{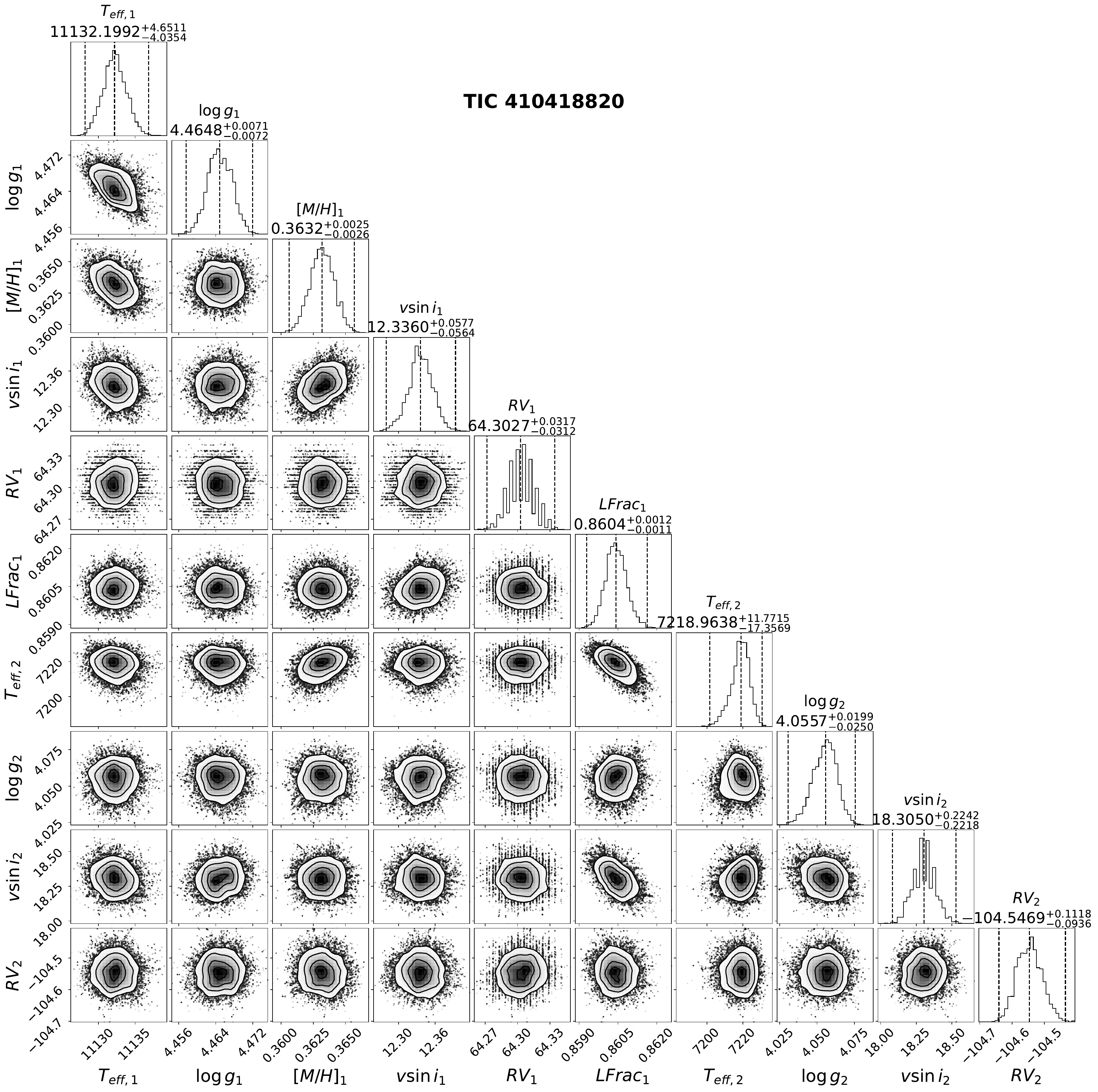}
\caption{Example corner plot showing posterior probability distribution for the atmospheric 
parameters from 5\,000 walkers simulations for TIC\,410418820, determined using the code. Contour 
levels correspond to 1, 2, and 3$\sigma$, and the histograms on the diagonal represent the 
posterior distribution for each parameter, with the mode and internal 99\% confidence levels 
indicated. More realistic errors are discussed in the text.
}
\label{mcmc_rob_e}
\end{figure*}
%
%

In order to generate synthetic spectra, the grid that is utilised is the pre-calculated 
grid that is included with the \isp\ software{\footnote{These grids and more can be 
		obtained from \url{https://lweb.cfa.harvard.edu/~sblancoc/iSpec/grid/}}}. Depending on the 
$T_{\rm eff}$, it might be necessary to transition between various grids when 
working with solutions. In our algorithm, generally, a spectrum grid is utilised for 
$T_{\rm eff}$ solutions that are less than 15\,000\,K. The \spectrumgrid\ was created 
using the \spectrum\ code \citep{1994AJ....107..742G}, ATLAS9.Castelli model atmospheres 
\citep{2003IAUS..210P.A20C}, and the line lists provided alongside the \spectrum\ 
code. These were adjusted to work with \isp\ software.

Forty-eight walkers were used to explore the posterior distributions for each solution 
to the systems. The walkers were initialised using randomly chosen beginning places as 
determined by the MCMC analysis. Subsequently, a burn-in solution was produced by 
employing MCMC techniques, with a minimum of 500 iterations per walker. Afterward, we 
advanced to the actual solution stage. This task was executed with a total of 5\,000 
iterations. The computing time may vary depending on the chosen grid and the temperature 
of the components. The optimisation uses the natural logarithm of the likelihood 
function, denoted as \[ \ln(p)=-0.5\sum(\frac {(y_m-y_o)^2}{\sigma_o^2}+\ln(2\pi\sigma_o^2)). \] The model 
data, observed data, and per-point uncertainty on the observed data are represented 
by $y_m$, $y_o$ and $\sigma$, respectively. The data sets with the highest signal-to-noise 
ratio were used for the MCMC analysis of all the examined systems. This enabled us to 
identify the posterior distributions inside the parameter space that closely align with 
our initial answer. Convergence of the MCMC runs may be assessed by analysing the 
modality of the posterior distribution's topology. A convergent MCMC run that is not 
influenced by the beginning conditions is extremely probable to have reached the global 
minimum \citep{2018ApJS..236...11H}. However, it is important to note that MCMC sampling 
cannot ensure the identification of the global minimum in the parameter space.

The process of MCMC involves constructing a reversible Markov chain that reaches an 
equilibrium distribution matching the target posterior distribution, allowing for the 
generation of samples. The posterior distributions offer valuable insights into the 
likelihood of a particular set of parameters accurately representing the gathered 
data. This research does not focus on the technical details of MCMC. Additional 
information on our code can be found in \citet{2023MNRAS.526.5987C}, as well.

Visualisation of the posterior distributions is accomplished by the use of extended 
corner plots, which exhibit cross-sections in two dimensions, as demonstrated. The 
illustration in Figure~\ref{mcmc_rob_e} is an illustration of a plan like this. It 
is challenging to determine if the MCMC walkers have thoroughly explored the posterior 
distribution. However, we can evaluate the progress of these MCMC runs using specific 
heuristic measures. According to \citet{2018ApJS..236...11H}, in order for an MCMC run 
to be judged as having achieved proper sampling, it is essential for each walker to 
have crossed the high-probability region of the parameter space numerous times. Indeed, 
\citet{2018ApJS..236...11H} defines the autocorrelation time per parameter as the number 
of steps necessary for the walkers to acquire independent samples of that parameter. This 
method determines the autocorrelation time for each parameter. The MCMC sampler is 
performing optimally when the chains show overlap and cover parameter values across 
the entire range of samples. 

Regarding the implementation of our spectrum analysis technique, Figure~\ref{mcmc_rob_e} 
demonstrates our solutions. With the UVES spectra, we are now able to provide accurate classifications 
for the target systems detailed in Table~\ref{tab:atmosphericParam}, taking into account the limitations 
set by the parameter errors.

%

%
\begin{table}
\caption{Stellar atmospheric parameters determined using the methods as described in detail in \S~\ref{spec_analysis}. In 
column two we list the T$\rm_{eff}$ determined from the Mg {\i} b triplet at 5183, 5172 and 5167\ \AA, as diagnostics of 
several important stellar parameters, such as the $T_{\rm eff}$, $\log g$ and [M/H]. For each system, the first 
parameter is for the primary component, while the second parameter is for secondary component.
}
\label{tab:atmosphericParam}
\centering
\begin{tabular}{lcccc}
\hline
\hline
Catalog~IDs                        &T$\rm_{eff}$          &$\log g$             &[M/H]                                  &$v\sin i_{\rm obs}$  \\
                                   &(K)                   &(dex)                &(dex)                                  &(km~s$^{-1}$)        \\
\hline
\multirow{2}{*}{KIC\,4932691}      &7\,580(450)  &4.16 [fixed]$^a$  &\multirow{2}{*}{-0.082(13)} &18(8)       \\
                                   &6\,935(1200) &4.20 [fixed]$^a$  &                                       &13(12)      \\ 
\multirow{2}{*}{TIC\,7695666}      &6\,280(300)           &4.32(7)     &\multirow{2}{*}{0.095(13)}             &9(1)                 \\
                                   &6\,050(350)           &4.11(10)    &                                       &9(1)                 \\ 
\multirow{2}{*}{TIC\,410418820}    &11\,100(250)          &4.47(4)     &\multirow{2}{*}{0.363(9)}              &12(1)                \\
                                   &7\,220(320)           &4.06(9)     &                                       &14(2)                \\ 
\multirow{2}{*}{EPIC\,203929178}   &6\,785(200)  &4.34(9)     &\multirow{2}{*}{0.035(20)}    &22(1)       \\
                                   &6\,777(300)  &3.49(9)     &                                       &29(1)       \\
\multirow{2}{*}{TIC\,213047427}    &7\,195(210)           &4.40(2)     &\multirow{2}{*}{0.231(19)}             &9(1)                 \\
                                   &6\,390(320)           &4.37(3)     &                                       &13(2)                \\
\hline
\end{tabular}\\
\begin{tablenotes}[]
\item Note: We adopted the same metallicity for both components.\\
$^a$ In the case of KIC\,4932691, we adopt the $\log g$ values computed from preliminary binary modelling. Due to the low SNR of APOGEE spectra we fixed $\log g$ values in atmospheric analysis to avoid convergence problem and parameter degeneration.
\end{tablenotes}   
\end{table}
%
%

%
%
%
%

\subsection{Binary Modelling and Absolute Dimensions}\label{pywd}
The light curves of target systems exhibit characteristic features of detached eclipsing binaries, evident 
in their shapes. The accurate determination of the beginning and ending moments of eclipses in 
binary systems is made possible by the presence of spherical or slightly ellipsoidal components 
in these systems. In addition, during the periods between eclipses, the level of light remains 
relatively stable or undergoes minimal fluctuations that are difficult to quantify using 
detectors. This can be attributed to factors such as reflection effects, slightly ellipsoidal 
components, or physical variations. In order to do a comprehensive analysis of the eclipse 
characteristics of this particular kind of star, we employ PyWD2015 
\citep{PyWD2015_2020CoSka..50..535G}, a Python framework that provides a graphical interface 
for the 2015 edition of the Wilson-Devinney (WD) algorithm \citep{WD_MAIN_1971ApJ,WD2015_2014ApJ}. This 
framework enables us to simultaneously modelling the light curves and radial velocities. PyWD2015 
provides a variety of useful features for visually monitoring the convergence of consecutive 
iterations, and it enables quick and efficient adjustment of solution techniques.

Our approximation involved maintaining a constant $T_{\rm eff}$ of the hotter primary 
component, which was calculated based on the atmospheric study of the systems discussed in 
the preceding section. The light curves clearly exhibit the expected characteristics of A-F 
stars, leaving no room for doubt regarding the temperatures. Shapes of light curves and our trial modelling attempts strongly indicate that all of our target systems are detached eclipsing binaries, thus we adopt operation mode of 2, which is proper for detached eclipsing binaries. We obtained the monochromatic 
($X,Y$) limb-darkening coefficients by interpolating from the tables of 
\citet{1993AJ....106.2096V}. The gravity-darkening exponents were set to standard values of $g_1$ and $g_2$ 
\citep{1967ZA.....65...89L,1969AcA....19..245R}, while the bolometric albedos were assigned as 
$A_1$ and $A_2$ (refer to Table~\ref{tab:salt_table}). In addition, both components were set to rotate synchronously 
with a rotation factor of 1.0, and the reflection treatment was implemented. For both components, we adopt coarse (NL) and fine (N) grid values as 15 and 30, respectively. The parameters that 
were adjusted included the orbital ephemeris ($T_0$,$P$), the system velocity ($\gamma$), the semi-major 
axis ($a$), the mass ratio ($q$), the orbital inclination ($i$), the surface temperature ($T_2$) of 
the secondary component, the dimensionless surface potentials ($\Omega_{1,2}$) of the components, and 
the monochromatic luminosity ($L_1$). Through further iteration, it has been found that the use of 
slightly inaccurate limb-darkening coefficients does not have a significant impact on the final 
errors of solution-based parameters for $K2$ and \tes\ data. Furthermore, we conducted tests on 
the third light ($l_3$), however we did not observe any discernible impact from these factors 
in the binary modelling. 

Table~\ref{tab:salt_table} presents the binary solutions for the best model enabling us to accurately 
define the absolute dimensions of the systems. The \jktabs\ code \citep{2005A&A...429..645S} was 
utilised to ascertain the physical characteristics of the systems. The code utilised input parameters 
$P$, $i$, $r_{1,2}$, $K_{1,2}$, and $T_{\rm eff1,2}$, along with their corresponding uncertainties 
listed in Table~\ref{tab:salt_table}. The results obtained from the code are also 
displayed in same table at the bottom. The resulting parameters were utilised to position 
the components on the HR diagrams in order to analyse their evolutionary status. The luminosity ($L$) and 
bolometric magnitudes ($M_{\rm bol}$) were derived by adopting $T\rm_{eff}$ = 5\,780\ K and $M\rm_{bol}$ = +4.73 
for solar values \citep{2000PhT....53j..77C}. 

The calculated velocities for the synchronised rotation of each component were $v_{\rm sync}$. Some 
eclipsing binaries have eccentric orbits. For individuals whose secondary light minima occur at 
around 0.5, we made the assumption that their eccentricities were equal to zero. This assumption 
is supported by evidence from short-period binaries, as they often exhibit circular orbits 
\citep{2016ApJ...829...34S}. In the case of long-period eccentric binaries, the occurrence of 
light minima at phase near 0.5 is only observed when the longitude of the periastron is close 
to 90$^{\circ}$. However, this scenario is unlikely to happen, as explained in equation 18 of 
\citet{2016AJ....151..139M}. We thoroughly examined the existing literature on individuals with 
noticeable eccentricities. Real synchronous rotation is unattainable in an eccentric orbit, hence 
a \textit{pseudo-synchronous} period $P_{\rm ps}$ is computed instead. In order to do this, we employed 
equation (42) as presented in \citet{1981A&A....99..126H}. The equation depicts the relationship 
between $P_{\rm ps}$, which stands for the period of the planet-star system, $e$, which represents 
the eccentricity of the orbit, and $P_{\rm orb}$, which signifies the orbital period. Pseudo-synchronous 
spin can result in a lack of overall torque during each orbit, preventing any changes in the spin 
\citep{1981A&A....99..126H}.

\jktabs\ can also estimate the distance to targets, using $T_{\rm eff}$ of two components, 
estimated metallicity, $E$(\B-\V) and apparent magnitudes via various calibrations. We used the interstellar 
reddening values $E$(\B-\V) to calculate the distance of the systems by using the remarkable approach 
to determine the average interstellar reddening values of $E$(\B-\V) is by analysing the Na {\sc i} 
(D$_2$ at 5889.951 \AA, D$_1$ at 5895.924 \AA)  spectral lines, which are prominent in the stellar 
spectrum \citep{1997A&A...318..269M}. The $E$(\B-\V) values that were measured are presented 
in Table~\ref{tab:salt_table}. The mean distance calculated for systems was consistent with all 
systems when utilising the trigonometric parallax from the $Gaia$ DR3 \citep{2023A&A...674A...1G}. 
%

%
%
\begin{figure*}
	\center
	\includegraphics[width=0.25\textwidth]{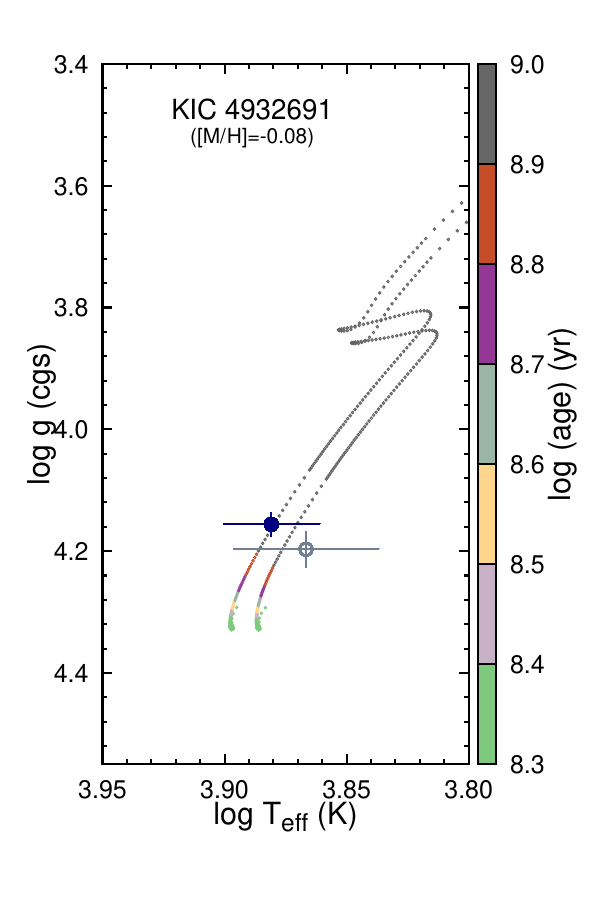}
	\includegraphics[width=0.25\textwidth]{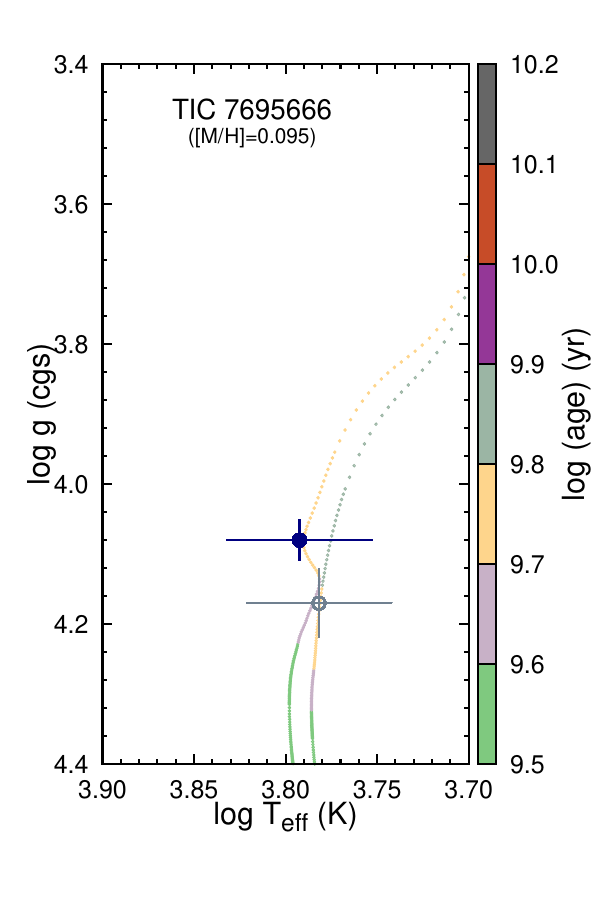}
	\includegraphics[width=0.25\textwidth]{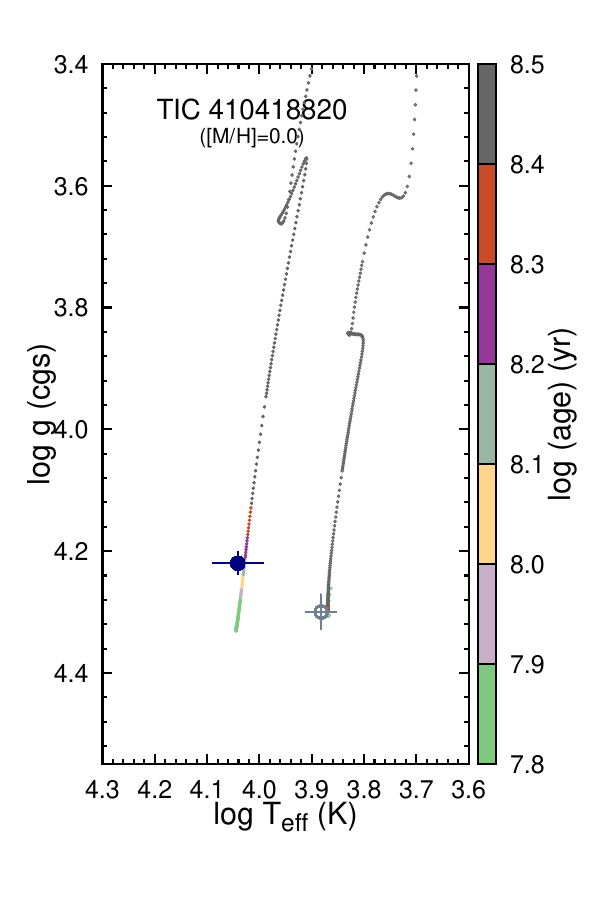}
	\includegraphics[width=0.25\textwidth]{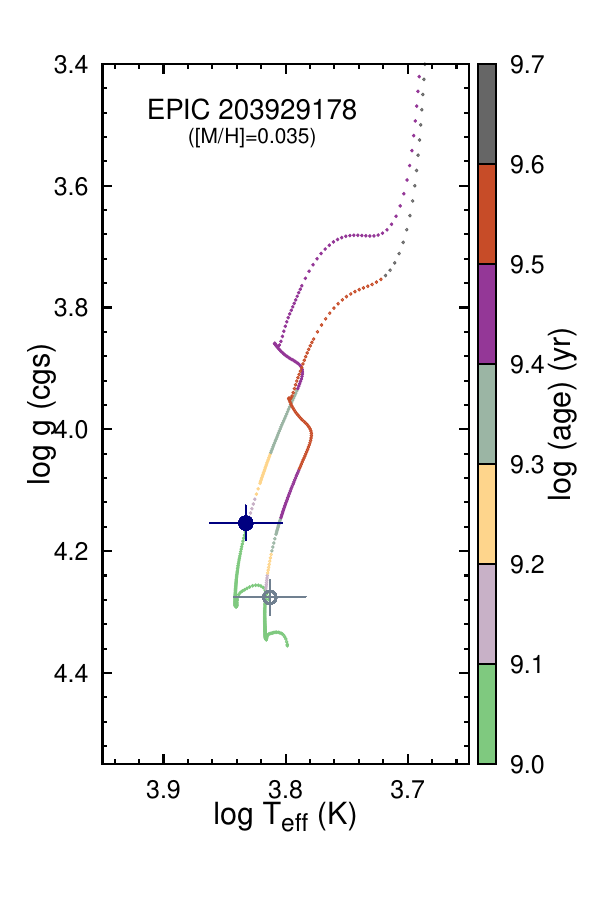}
	\includegraphics[width=0.25\textwidth]{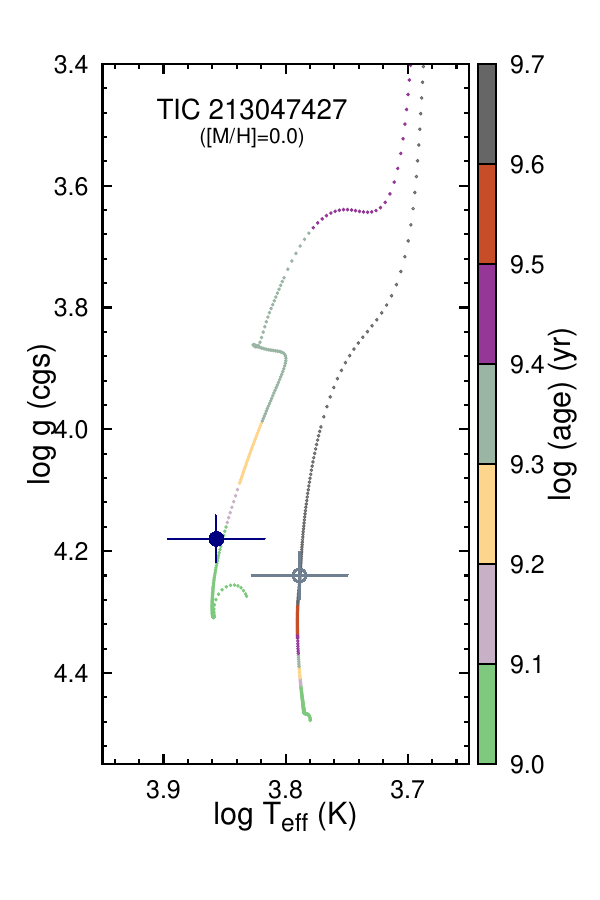}
	\caption{Comparison of our results with MESA evolution tracks on $T_{\rm eff}$/log $g$ planes. Primaries are 
		shown with filled circle symbols, and secondaries with open circle symbols, both with error bars from 
		Table\,\ref{tab:salt_table}. The coloured points are the evolutionary tracks that best match to calculated 
		component masses and the colours indicate the stellar age with the colour bars at the right side of the plots.}
	\label{iso_hr}
\end{figure*}
%
%

\section{Evolutionary status}\label{sec_iso}
Studying the location and extent of the observed and predicted instability strips is an essential test for 
the models. Figure\,\ref{iso_hr} presents a comparison between our analysis results and the theoretical 
MESA (Modules for Experiments in Stellar Astrophysics) 
evolution tracks created as a component of the MESA Isochrones and Stellar Tracks 
project \citep[MIST\footnote{\url{http://waps.cfa.harvard.edu/MIST/interp_isos.html}} 
v1.2,][]{2016ApJS..222....8D, 2016ApJ...823..102C}. We determine the age at which the observed or computed 
properties of both components most accurately represent the entire system. We choose $T_{\rm eff}$ - $\log g$ 
planes to compare our data as the temperature, $T_{\rm eff}$, of a star is a highly reliable parameter 
obtained by spectral research. A key point to note is that the measured $T_{\rm eff}$ values of the 
components are independent parameters.

The MIST models essentially include rotation but do not integrate the MESA binary module, which has the 
capability to evolve the structure of two stars \citep{2015ApJS..220...15P}. Nevertheless, the variations 
in stellar parameters between the main sequence, early RGB phases, and masses below 2 $M_{\odot}$ are 
negligible compared to our margin of error in measurements, and they do not have a substantial impact on 
the estimated ages. On the other hand, the pulsation could have a significant impact on the radii and 
effective temperatures of low-mass and sub-giant stars, making it particularly relevant for some of our 
objects. The most recent isochrones for stars are able to reproduce precisely those properties. Based on 
our findings, it seems that the pulsating level may not have an impact on our age determination in these 
cases. The grid of isochrones was generated for a wide range of [Fe/H] abundance values and ages, covering 
a logarithmic scale. The [Fe/H] abundance values ranged from -4.0 to 0.5\,dex with 0.05\,dex steps, while 
the ages ranged from 10$^{7}$ to 10$^{10.5}$\,yr. The isochrones provided are an accurate match to the 
properties of systems, although not excellent.

\begin{landscape}
	\begin{table} 
		\centering
\caption{Results for  $\gamma$ Dor stars in eclipsing binaries. Uncertainties of the parameters are given in parenthesis for the last digits.}
\label{tab:salt_table} 
\begin{tabular}{lcccccccccc}
\hline
Parameter                                      & \multicolumn{2}{c}{KIC\,4932691}                    & \multicolumn{2}{c}{TIC\,7695666}  & \multicolumn{2}{c}{TIC\,410418820} & \multicolumn{2}{c}{EPIC\,203929178}                   & \multicolumn{2}{c}{TIC\,213047427} \\
                                               &    $primary$    &   $secondary$                     &    $primary$    &   $secondary$   &    $primary$     &   $secondary$   &    $primary$    &    $secondary$                      &    $primary$    &   $secondary$    \\ \hline
T$_0$(BJD$-2,400,000$)$^a$                     & \multicolumn{2}{c}{54972.5106(5)}          & \multicolumn{2}{c}{58429.2431(1)} & \multicolumn{2}{c}{58633.4290(2)}  &  \multicolumn{2}{c}{56895.8866(2)}           & \multicolumn{2}{c}{58455.3931(1)}  \\ %
P (day)                                        & \multicolumn{2}{c}{18.112104(2)}                    & \multicolumn{2}{c}{17.554631(4)}  &  \multicolumn{2}{c}{8.569026(3)}   &   \multicolumn{2}{c}{2.307883(16)}           &  \multicolumn{2}{c}{8.795873(2)}   \\
$a$ $sin\,i$ (R$_{\odot}$)                     &  \multicolumn{2}{c}{42.57(21)}             &  \multicolumn{2}{c}{37.384(88)}   &   \multicolumn{2}{c}{28.145(22)}   &   \multicolumn{2}{c}{10.150(16)}             &   \multicolumn{2}{c}{24.651(19)}   \\ %
$\gamma$ (kms$^{-1}$)                          &    \multicolumn{2}{c}{88.5(7)}             &    \multicolumn{2}{c}{25.3(4)}    &     \multicolumn{2}{c}{1.1(2)}     &    \multicolumn{2}{c}{-20.0(4)}              &    \multicolumn{2}{c}{20.6(2)}     \\ %
K$_{1,2}$ (km s$^{-1}$)                        & 64.0(5)&65.3(4)                   &     62.9(2)     &     67.3(3)     &     62.4(4)      &    103.9(7)     &    105.6(2)     &     117.0(3)      &     60.7(3)     &     80.9(4)      \\ %
$e$                                            &   \multicolumn{2}{c}{0.392(2)}             &   \multicolumn{2}{c}{0.561(9)}    &    \multicolumn{2}{c}{0.027(3)}    &    \multicolumn{2}{c}{0.000(1)}                       &    \multicolumn{2}{c}{0.000(2)}    \\ %
$\omega$(${\degr}$)                            &    \multicolumn{2}{c}{163(2)}              &    \multicolumn{2}{c}{176(2)}     &     \multicolumn{2}{c}{300(3)}     &      \multicolumn{2}{c}{1(1)}                         &      \multicolumn{2}{c}{1(2)}      \\ %
$q$ (M$_2$/M$_1$)                              &   \multicolumn{2}{c}{0.980(10)}            &   \multicolumn{2}{c}{0.935(5)}    &    \multicolumn{2}{c}{0.601(6)}    &    \multicolumn{2}{c}{0.903(4)}              &    \multicolumn{2}{c}{0.750(3)}    \\ %
$i$ (${\degr}$)                                &    \multicolumn{2}{c}{86.0(1)}             &    \multicolumn{2}{c}{88.3(4)}    &    \multicolumn{2}{c}{88.7(2)}     &    \multicolumn{2}{c}{77.4(5)}              &    \multicolumn{2}{c}{86.6(2)}     \\ %
T$_{\rm eff\,1,2}$ (K)                         & 7\,580$^b$[fixed] &   7\,356(120)   & 6\,280$^b$[fixed] &   6\,170(120)   & 11\,100$^b$[fixed] &   7\,620(290)   & 6\,785$^b$[fixed] &    6\,492(150)    & 7\,195$^b$[fixed] &   6\,200(150)    \\ %
$\Omega_{1,2}$                                 &   25.967(66)    &   26.923(61)    &   27.486(33)    &   24.632(33)    &    14.382(41)    &   12.883(44)    &    7.058(29)    &     7.774(52)     &   20.261(44)    &    17.288(45)    \\ %
$A_{1,2}$                                      &   \multicolumn{2}{c}{1.0[fixed]}             &   \multicolumn{2}{c}{1.0[fixed]}    &    \multicolumn{2}{c}{0.5[fixed]}    &    1.000(38) & 0.5[fixed]             &    \multicolumn{2}{c}{1.0[fixed]}    \\ %
$g_{1,2}$                                      &   \multicolumn{2}{c}{1.0[fixed]}             &   \multicolumn{2}{c}{1.0[fixed]}    &   \multicolumn{2}{c}{0.32[fixed]}    &    \multicolumn{2}{c}{0.32[fixed]}                      &    \multicolumn{2}{c}{1.0[fixed]}    \\ %
$X_{1,2}^c$                                    &      0.441      &      0.445      &      0.337      &      0.292      &      0.352       &      0.307      &      0.473      &       0.490       &      0.299      &      0.299       \\ %
$l_{1}/(l_1 + l_2)$                            &    \multicolumn{2}{c}{0.56(1)}             &    \multicolumn{2}{c}{0.49(5)}    &    \multicolumn{2}{c}{0.82(2)}     &     \multicolumn{2}{c}{0.64(1)}              &    \multicolumn{2}{c}{0.63(2)}     \\ %
$l_{3}/(l_1 + l_2 + l_3)$                      &      \multicolumn{2}{c}{---}                        &      \multicolumn{2}{c}{---}      &      \multicolumn{2}{c}{---}       &    \multicolumn{2}{c}{0.116(17)}             &      \multicolumn{2}{c}{---}       \\ %
$r_{\rm 1,2}$                                  &    0.0411(1)    &    0.0388(1)    &    0.0395(2)    &    0.0422(3)    &    0.0725(1)     &    0.0519(1)    &    0.1629(1)    &     0.1345(1)     &    0.0619(1)    &    0.0566(1)     \\ %
$\Sigma$ $W(O-C)^2$                            &    \multicolumn{2}{c}{0.0005}              &    \multicolumn{2}{c}{0.0031}     &     \multicolumn{2}{c}{0.0033}     &     \multicolumn{2}{c}{0.0006}               &     \multicolumn{2}{c}{0.0033}     \\ \hline
Absolute parameters                            &                 &                                   &                 &                 &                  &                 &                 &                                     &                 &                  \\ \hline
M$_{1,2}$ (M$_{\odot}$)                        &    1.597(24)    &    1.565(27)    &    1.179(15)    &    1.102(25)    &    2.550(19)     &    1.532(27)    &    1.492(8)     &    1.346(8)       &    1.486(23)    &    1.115(25)     \\ %
R$_{1,2}$ (R$_{\odot}$)                        &    1.750(10)    &    1.652(9)     &    1.478(35)    &    1.578(45)    &    2.041(35)     &    1.461(45)    &    1.694(3)     &     1.399(3)      &    1.536(32)    &    1.424(44)     \\ %
$\log(g_{1,2})$ (cgs)                          &    4.155(3)     &    4.196(4)     &    4.170(6)     &    4.084(6)     &     4.225(7)     &    4.294(9)     &    4.154(2)     &     4.276(2)      &    4.183(11)    &    4.244(15)     \\ %
$(v_{1,2}\sin i)_{\rm calc}$ (km s$^{-1}$)$^d$ &      5(1)       &      5(1)       &      6(1)       &      6(2)       &      12(1)       &      9(2)       &      37(1)      &       31(1)       &      8(1)       &       8(2)       \\ %
$\log(L_{1,2}/L_{\odot})$                      &    0.960(35)    &    0.858(59)    &    0.486(11)    &    0.513(15)    &    1.761(25)     &    0.812(33)    &    0.739(64)    &     0.496(67)     &    0.756(31)    &    0.431(21)     \\ %
M$_{\rm bol\,1,2}$ (mag)                       &    2.350(87)    &    2.605(148)   &   3.534(112)    &   3.468(117)    &    0.348(115)    &   2.719(124)    &   2.901(161)    &    3.509(167)     &   4.159(119)    &    4.423(113)    \\ %
(m--M)$_V$ (mag)                               &  \multicolumn{2}{c}{11.336(82)}            &   \multicolumn{2}{c}{8.104(40)}   &   \multicolumn{2}{c}{5.505(50)}    &    \multicolumn{2}{c}{8.459(85)}            &   \multicolumn{2}{c}{6.133(77)}    \\ %
$E(B-V)$ (mag)                                 &   \multicolumn{2}{c}{0.204(11)}            &   \multicolumn{2}{c}{0.111(9)}    &   \multicolumn{2}{c}{0.090(12)}    &    \multicolumn{2}{c}{0.267(37)}             &    \multicolumn{2}{c}{0.021(4)}    \\ %
$d$ (pc)$^e$                                   &  \multicolumn{2}{c}{1\,930(9)}            &    \multicolumn{2}{c}{356(7)}     &     \multicolumn{2}{c}{111(3)}     &     \multicolumn{2}{c}{617(11)}           &     \multicolumn{2}{c}{169(3)}     \\ %
$d$ (pc)$^f$                                   &  \multicolumn{2}{c}{1\,965(45)}                     &    \multicolumn{2}{c}{368(3)}     &     \multicolumn{2}{c}{124(1)}     &     \multicolumn{2}{c}{677(86)}                       &     \multicolumn{2}{c}{156(1)}     \\ 
$\log (Age)$ (yr)& 8.96(4)&8.91(8)&9.71(1)&9.78(3)&8.18(7)&8.28(39)&9.12(7)&9.05(17)&9.05(9)&9.67(5)\\
\hline
\end{tabular} 
\begin{tablenotes}
\item \emph{Note}: 
$^a$Mid--time of the primary eclipse. $^b$T$_{\rm eff\,_1}$ were found in atmospheric 
analysis in \S~\ref{spec_analysis}. $^c$$X$ denotes linear coefficients of limb darkening. $^d$Theoretical 
projected rotational velocity calculated by \jktabs\ code under the assumption of (pseudo) synchronous 
rotation. $^e$The \jktabs\ distance is calculated from 2MASS magnitudes. $^f$From $Gaia$ DR3 parallaxes \citep{2023A&A...674A...1G}.
		\end{tablenotes}
	\end{table} 
\end{landscape}

Figure\,\ref{iso_hr} illustrates the position of the $\gamma$\,Dor type pulsating components within 
the HR diagram, highlighting the positions of the zero-age main sequence (ZAMS) and the terminal-age 
main sequence (TAMS). The pulsating components usually spread along the age range of 0.1 to 3.2\,Gyr. The 
roughly estimation of the ages is presented in Table\,\ref{tab:salt_table} for each system. Observations 
indicate that these stars generally cover a region of the HR diagram characterised by temperatures ranging 
from approximately 7\,200 to 7\,550 K on the ZAMS and about 6\,900 to 7\,400 K near the terminal-age main 
sequence \citep{2002MNRAS.333..251H}, where they exhibit pulsations in gravity modes. Up to this point, these 
boundaries have been established solely through observation. In this study, our samples match with the observational 
address. Precise measurements of the masses and ages of main sequence stars, along with evolutionary models that 
connect these quantities to observable characteristics, are crucial for addressing numerous questions in the study 
of pulsating stars and stellar evolution.
While there is a general consistency in the evolutionary tracks, the 
two systems do not align with theoretical evolutionary models: TIC\,410418820 and TIC\,213047427. Kiel diagrams suggest that both systems should have solar metallicity (see Fig.\,\ref{iso_hr}), while spectroscopic analysis indicates significantly metal-rich components. That difference may partly originate from uncertainties in normalization level of analysed spectra. However, we have no certain explanation for the difference, which remains as an open question.

%
%
%
\begin{figure*}
	\center
\includegraphics[width=1.025\textwidth]{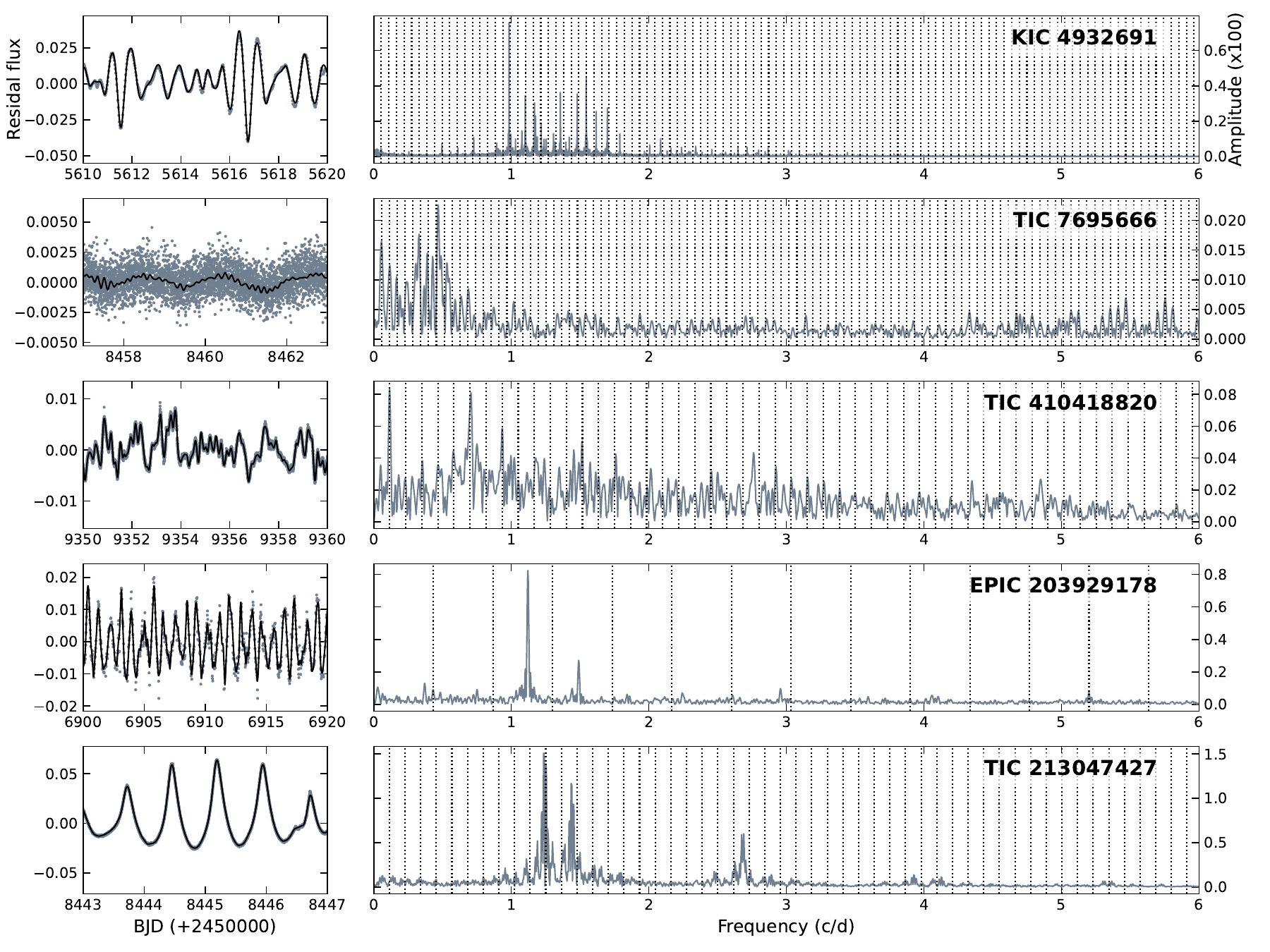}
\caption{Pulsation spectra of five $\gamma$ Dor stars observed with \kep\ and \tes. The stars are arranged in a 
certain order based on the information provided in Table \ref{tab:salt_table}, which includes their 
orbital values and other absolute properties. The cadence residuals for each target are computed 
and presented in the left panels for each star, based on the best-fitting light curve models. This 
is done within a restricted range to ensure that the fit is visible and to assist in guiding the eye. The 
fits in various regions are calculated based on the all-frequency values that are fitted to the data. The 
amplitude spectrum of all long-cadence residuals is displayed in the panel on the right. The zoomed 
view shows the amplitude spectrum spanning from 0 to 6 cycles per day. The vertical dashed line 
in each panel represents the orbital frequency and its harmonics of the respective system. The graphic accurately 
depicts the primary genuine frequencies of each star, disregarding the orbital frequency.
	}
	\label{freq_fig}
\end{figure*}
%
%

\section{Light residuals, pulsational characteristics and correlations}\label{puls}
Apart from the eclipses, the light curves of the systems show clear additional variability that mainly can 
be attributed to pulsations. For each target, we subtracted our best-fitting light curve model from the 
observed data to obtain precise residuals. Since residuals data do not contain any binarity effect 
(like reflection) pulsation signals can be investigated safely. In order to extract frequency signals 
from the residuals, we apply a multi-frequency analysis by using \sigspec\ software \citep{2007A&A...467.1353R} 
to the outside eclipse data. \sigspec\ calculates the significance spectrum for a given 
non-equally spaced time series data and evaluates the probability density function of discrete Fourier 
transform (DFT) peaks. Multi-frequency signals can be extracted iteratively from data via a pre-whitening 
procedure. In each step of the iterative pre-whitening, \sigspec\ calculates the frequency, amplitude, phase, 
and significance level of the highest peak (signal) in DFT. Then, removing the signal from the time-series 
data, the process is repeated. This iterative process runs until the significance level is below a defined 
limit. The analysis with \sigspec\ yielded formally significant frequencies with $S/N$ values higher than 5.0 
(Spectral Significance parameter; $\sigma$ = 5.5 proposed by \citet{2007A&A...467.1353R} as theoretically 
equivalent of $S/N$ = 4.0). Applying the significance thresholds given above, \sigspec\ finds formally 
significant frequencies for the light curve residual data. Table\,\ref{tab:Freqs} contains the unique 
frequencies found for the $\gamma$ Dor systems. $\gamma$\,Dor stars show combination frequencies, due 
to the nonlinear effects of their observational data, we search for those combinations by calculating with 
$|f_i - (nf_i + mf_k)| < \epsilon$, where $n$ and $m$ are integers (1,2,3), and $\epsilon$ is the Rayleigh 
resolution. If the difference between two frequencies is less than the Rayleigh resolution, the two frequencies 
are indistinguishable \citep{2012AN....333.1053P}; and if a frequency could be combined with two main frequencies 
with larger amplitudes, it is omitted in Table\,\ref{tab:Freqs} last column. 

The errors of all values listed in Table\,\ref{tab:Freqs} are derived according to 
\citet{2008A&A...481..571K}. The light residuals after removing the binary effects from the observed 
data are plotted in Figure\,\ref{freq_fig} as $\rm mmag$ versus BJD, where the left panels present a short 
section of the residuals. The amplitude spectrum of the residuals is shown in the right panel of 
Figure\,\ref{freq_fig}, displaying significant frequencies and orbital harmonics represented by dashed lines.
%

%
%
\begin{figure*}
\center
\includegraphics[width=0.87\textwidth]{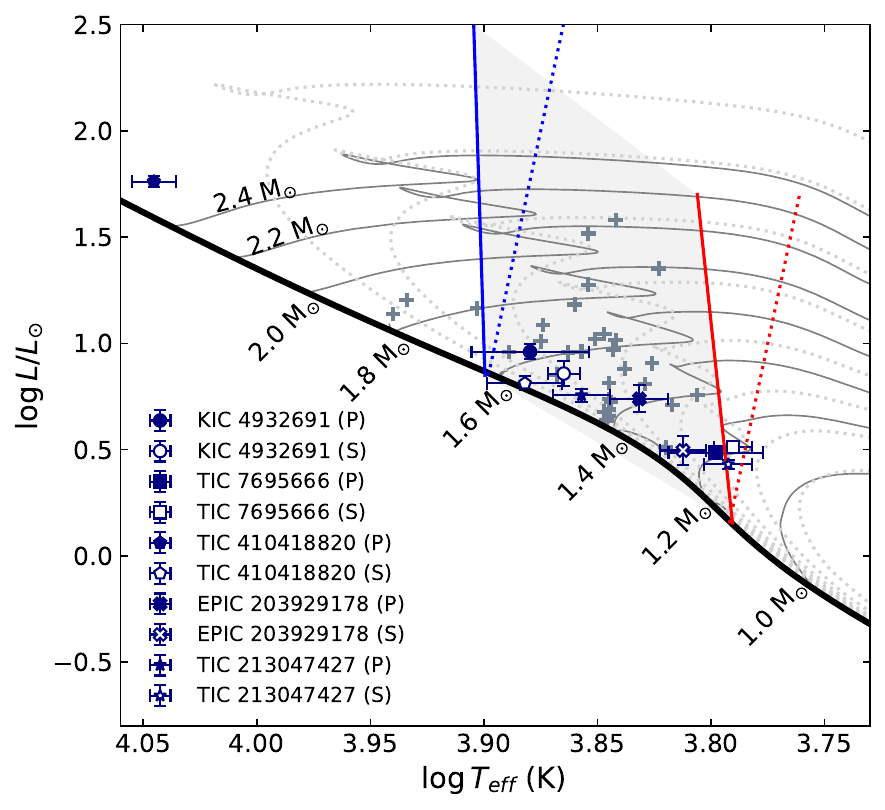}
\caption{Positions of the $\gamma$\,Dor stars analysed in this study (represented by different symbols) on 
the HR diagram alongside all 34 verified $\gamma$\,Dor stars \citep[displayed with plus symbols,][]{2020RMxAA..56..321O}. The 
calculated instability strip, represented by the solid straight lines and gray-colored zone are being compared to the 
observational $\gamma$\,Dor instability strip, shown by the dotted straight lines \citep[][]{2002MNRAS.333..251H}. The 
estimated zero-age main sequence \citep{1996MNRAS.281..257T}, represented by the solid line.
}
	\label{hr}
\end{figure*}
%
%

\citet{2018NewA...62...70I} introduced the comprehensive catalog of $\gamma$\,Dor pulsators in 
eclipsing binaries with known parameters for the first time, and it listed 24 systems confirmed 
as binary spectroscopically and photometrically. With the inclusion of the ten stars from the 
literature and the five stars from this study, the total count reached 39. Note that 
\citet{2018NewA...62...70I} did not include the only photometrically studied $\gamma$\,Dor 
candidates from the literature in this catalog, so the total number of 39 samples in this 
study may look small; however, the systems studied are obtained from reliable observation 
analysis with well-known absolute parameters.

From a pulsational point of view, there is a clear distinction between their position on the HR 
diagram because their pulsational behavior is currently known to the 
$\gamma$\,Dor type pulsating stars and its detailed analysis contributes to the sample of detached 
binaries with a $\gamma$\,Dor member. The distinction is even clearer if one considers the known 
locations of such systems on the HR diagram given by \cite{2020RMxAA..56..321O}. Using best 
fitting spectroscopy+LC models, we plot log $T_{\rm eff}$ - log $L/L{_\odot}$ for our samples in 
Figure\,\ref{hr}.

By considering the values used in Figure\,\ref{hr}, some of the stars in our sample (both components 
of KIC\,4932691) fall outside the region given for $\gamma$\,Dor-type pulsating systems 
\citep{2020RMxAA..56..321O}. However, the distribution of the stars mostly confirm that the stars 
are indeed significantly unusual than the region of the known $\gamma$\,Dor stars. However, in the 
same region of the HR diagram where the $\delta$\,Sct and $\gamma$\,Dor instability strips overlap, 
$\gamma$\,Dor oscillations are also predicted to occur for $\delta$\,Sct stars 
\citep{1999A&A...351..582H,2002A&A...395..563S}. Apart from the single $\gamma$\,Dor type candidates 
hotter than the granulation boundary that \citet{2016MNRAS.460.1318B} has reported, so far, no 
detections of detached binaries with $\gamma$\,Dor-like oscillations have been reported, which 
is very close to the granulation boundary \citep{1989ApJ...341..421G}.

Figure\,\ref{hr} shows that our constructed $\gamma$\,Dor instability strip aligns closely with 
the observational strip from \citet{2016MNRAS.457.3163X}. Despite their compatibility, some 
sytems seem to exist outside the instability strip. The growing number of these stars suggests 
a shift in the boundaries of the instability region, despite the initial assumption that they are 
$\gamma$\,Dor \citep{2016MNRAS.460.1318B}. Keep in mind that these stars are pure $\gamma$\,Dor 
stars, and their parameters have been meticulously determined through spectral and photometric 
methods. This work emphasises a significant learning opportunity: by analysing the distribution 
of pure $\gamma$\,Dor, it is feasible to see and calculate a more accurate instability strip. The 
dotted and solid lines in Figure\,\ref{hr} represent the new blue and red boundaries of the instability 
strip for all observed $\gamma$\,Dor systems. With the advent of new pure $\gamma$\,Dor observations, these 
edges might be changed. The new instability strip, which is represented by the blue edge;
\[
\log T_{\rm eff} = 3.895-0.020 \times \left( \log \frac{\rm L} {\rm L_{\odot}}-1.0 \right)
,\] and the red edge; 
\[
\log T_{\rm eff} = 3.775-0.020 \times \left( \log \frac{\rm L} {\rm L_{\odot}}-1.0 \right)
,\] were both estimated with this work.

The influence of absolute parameters on the pulsational behavior is one of the 
major questions on this particular topic of hosting a $\gamma$\,Dor member's 
asteroseismology. There has been several discussion about it. \citet{2016NewA...45...36C} 
announced the first study which claims that there is an exact relation between pulsation 
and orbital periods that can be possibly increased with the samples. \citet{2018NewA...62...70I} 
updated a catalog including 23 cases and improved correlations between fundamental 
parameters. They noticed the first correlation between orbital ($P_{\rm orb}$) and dominant 
pulsation ($P_{\rm puls}$) periods for systems hosting a $\gamma$\,Dor member. This catalog 
is fairly considered a benchmark for asteroseismology, although the first correlation 
findings were given for eclipsing binaries hosting a $\gamma$\,Dor component. A few 
years later \citet{2020RMxAA..56..321O} and \citet{2020MNRAS.491.5980H} published the 
catalog with new samples, which also includes updated correlations between fundamental 
parameters according to the location of the stars on the HR diagram and the accuracy 
of their absolute parameters. 
%

%
%
\begin{figure*}
	\center
	\includegraphics[width=0.99\textwidth]{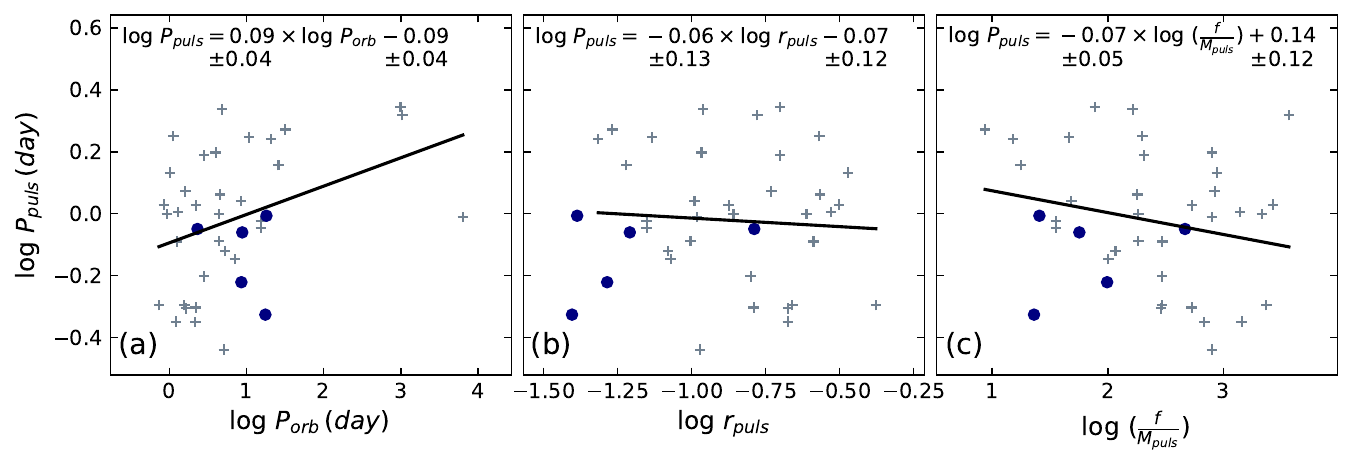}
	\caption{Relations between pulsation period $P_{\rm puls}$ and $P_{\rm orb}$, $r_{\rm puls}$, $F/M_{\rm puls}$ on logarithmic 
		scale. Blue, filled circle symbols denote the new $\gamma$\,Dor components of studied detached 
		systems. Filled plus are taken from \citet{2020MNRAS.491.5980H} and references therein. Dashed lines 
		show relations found by \citet{2016NewA...45...36C} while continuous lines are relations updated in 
		this study. Linear fit parameters are given in each window with uncertainties for the last digits.}
	\label{relat}
\end{figure*}
%
%

The $\gamma$\,Dor type pulsating components of the systems in this  study were placed among others of similar 
properties in the orbital-pulsational period relations, which is shown in Figure\,\ref{relat}. We demonstrated 
that the relation between pulsation period ($P_{\rm puls}$) and orbital period ($P_{\rm orb}$), the force per gram 
exerted on the surface of the pulsating star ($F/M_{\rm puls}$) and the fractional radius of the pulsating 
star ($r_{\rm puls}$). $F/M_{\rm puls}$, which is given by $\frac{F}{M_{\rm puls}}$=$\frac{G M_s}{(a-R_{\rm puls})^2}$, where $G$ 
is the Newton’s gravitational constant, $M_s$ is the mass of the secondary component, $a$ is the semi-major 
axis of the system, $M_{\rm puls}$ and $R_{\rm puls}$ are the mass and the radius of the pulsating 
component, respectively \citep{2016NewA...45...36C}. Even though the present analysis for these systems is 
significantly more detailed than the preliminary one of \citet{2020RMxAA..56..321O}, the new derived 
dominant frequencies are close enough to the previous values and tend to support the overall 
variations. However, given that these systems are the $\sim$15\,\% of the total available sample of 
eclipsing binaries with $\gamma$\,Dor component, the current correlation \citep{2020RMxAA..56..321O} tends to 
remain the same as before. 

The primary of KIC\,4932691 is among the first five $\gamma$\,Dor type stars with the longest $P_{\rm orb}$ values 
and is very close to the empirical fit. The pulsators of the other five systems follow well the empirical fit 
and the distribution of the other stars. On the other hand, the $\gamma$\,Dor component of EPIC\,203929178 is the 
remarkable pulsator of the sample but is also one of the most deviated cases from the fit. However, the deviation 
of the pulsator does not affect the overall variation of the change.

In Figure\,\ref{relat} (middle) the locations of the $\gamma$\,Dor components of the studied systems and other $\gamma$\,Dor 
type stars within the $\log r$\,-\,$\log P_{\rm puls}$ diagram is also shown. Considering that all of these systems are 
detached, contrary to the previous empirical fit, in this case, the contribution of the systems to a new empirical 
relation between $\log r$\,-\,$\log P_{\rm puls}$ is noticeable 
. The previous 
fit \citep{2020RMxAA..56..321O} included all the known systems with $\gamma$\,Dor components, whose absolute 
parameters are known, regardless of their Roche geometry. The differences between the two studies' correlations 
are not negligible. Both the slope and the correlation values have large deviations from each other, indicating 
that absolute parameters play a significant role in the pulsation frequencies. However, in the present study and 
for the following empirical correlation, only the detached eclipsing binaries with $\gamma$\,Dor components are 
taken into account. The five studied systems consist the $\sim$13\,\% of the whole sample of $\gamma$\,Dor stars 
(39 in total) and they contribute for the first time in this fit with updated and more accurate absolute 
parameters (Table\,\ref{tab:salt_table}). Therefore, the new empirical correlation between $\log r$\,-\,$\log P_{\rm puls}$ 
for the detached eclipsing binaries with a $\gamma$\,Dor component is given and is drawn in Figure\,\ref{relat} 
(middle) as well. The graph shows the distribution of the $\gamma$\,Dor stars in detached binaries and their 
corresponding correlations show that the proximity of the companion star is very important to 
decreasing of pulsation periods. The pulsational periods are decreasing while the surface gravities 
are increasing, similar to the $\delta$\,Sct stars \citep{2012MNRAS.422.1250L,2006MNRAS.366.1289S} and to those 
all radially pulsating variable stars. While the slope is about -3.3 for the $\delta$\,Sct  
\citep{2012MNRAS.422.1250L,2006MNRAS.366.1289S}, it is 
-0.88 \citep{2018NewA...62...70I} for the radially pulsating variable stars. The pulsation periods of $\gamma$\,Dor 
type pulsators seem to depend on the surface gravity similar to those radially pulsating variable stars proposed 
by \citet{1995AJ....110.2361F}. More studies of similar systems will enrich the sample of $\gamma$\,Dor stars with 
accurately determined absolute properties, which in turn, will improve knowledge of how interactions of components 
in close binary systems influence pulsations.

What can be clearly seen in Figure\,\ref{relat} (right) is the dominance of the pulsational periods of the oscillating 
primary components against the gravitational pull exerted per gram of the matter on the surface of the pulsators by 
the less massive companions. EPIC\,203929178 with the smallest orbital period among the five stars studied in this paper appears 
to have a shorter pulsational period with respect to the gravitational pull of the companion. The correlation 
coefficient indicates that the empirical relationship between pulsational periods of the oscillating primary 
components and the gravitational pull exerted per gram of the matter on the surface of the pulsators is 
clear (with a correlation coefficient of 0.41). As the gravitational pull exerted on the per gram matter of the pulsator is increased 
the pulsational period is decreased, confirming the empirical relationship found. The empirical relationship 
should, of course, be checked by increasing the number of samples. 

Correlations are usually worked out on the basis of samples. Each scatter plot in Figure\,\ref{relat} 
represents a dataset with two quantitative variables, and we are interested in the statistical 
relationships between these variables. At the left of Figure\,\ref{relat}, the variables have a positive 
correlation (with a correlation coefﬁcient of 0.36). It seems from the samples that binarity has a slight impact on the pulsation periods for orbital 
periods that are less than $\sim$13 days. The true change takes place when the $\gamma$\,Dor-type stars 
that were investigated in \citet{2022AJ....163..180H} are taken into consideration. As soon as we incorporate 
the long orbital period $\gamma$\,Dor stars that \citet{2022AJ....163..180H} investigated into the 
graph, the difference becomes readily apparent. It is unclear whether this is directly the consequence of 
tidal interaction has played a role in this. The significant amount of scattering that occurs around the 
$P_{\rm orb}$ - $P_{\rm puls}$ relation calls into question its utility; these targets must to be examined with the 
subsequent data to see whether or not more information can be acquired. Figure\,\ref{relat}'s 
middle and right graphs, on the other hand, do not make use of the three stars that were identified by 
\citet{2022AJ....163..180H} research because they are devoid of any light curve solutions. Also Figure\,\ref{relat}'s 
middle and right show a negative correlation (with a correlation coefﬁcient of -0.07 and -0.22, respectively).
%
%
The orbital configuration plays a role since various forms of tidal interactions are found in short- and 
long-period, circular $vs.$ eccentric systems. Tides can generate stellar pulsations, e.g. by a resonance 
mechanism such as in the eccentric 'heartbeat' systems \citep{2012MNRAS.420.3126F,2020ApJ...903..122C}. In 
turn, non-radial oscillations generated by tides may alter the evolution of close binaries 
\citep{2021RvMP...93a5001A}.

\section{Conclusions}\label{conc}
The current study establishes a quantitative framework for a comprehensive and in-depth examination of five 
eclipsing binaries with pulsating components. We were able to do a thorough statistical analysis utilising 
the available combination of \kep\ and \tes\ photometry and ground-based spectroscopy. The spectral categorization 
of its primary components, based on our spectroscopic data, allowed for reliable light curve studies and 
estimations of the absolute parameters and evolutionary phases of both eclipsing binary components. In 
addition to identifying the primary component as a $\gamma$\,Dor star, its pulsational features were 
precisely calculated. Even though our samples only have $\gamma$\,Dor pulsators, these are interesting 
for a number of reasons. Initially, we established that all of these stars are binary and possess a 
$\gamma$\,Dor pulsator. Furthermore, we have determined that all observed $\gamma$\,Dor fluctuations 
stem only from the primary component.

Additionally, we have made updates to the empirical instability strip for $\gamma$\,Dors, which now 
includes the $\gamma$\,Dors examined in this study. We carefully examined photometric and spectroscopic 
data in order to precisely mimic the blue and red edges of pulsating $\gamma$\,Dors. To achieve consistent 
stability for the established instability strip of $\gamma$\,Dor stars, some adjustment of parameters 
would be necessary, based on the initial theoretical $\gamma$\,Dor instability strip derived from 
\citet{2003ApJ...593.1049W} models. They either fall in between different sets or have varying 
slopes. Although we are not focusing on fine-tuning in this work, our empirical data could potentially 
be used in the future to confirm the accuracy of theoretical models. Undoubtedly, the validation would 
be highly appreciated and valuable in utilising the observations for comparing with models in situations 
where the quantity of pure $\gamma$\,Dors are insufficient to empirically ascertain the boundaries of the 
instability strip.

It has been stated in the past that pulsation period relations exist for $\gamma$\,Doradus variables in 
eclipsing binaries, along with a theoretical foundation for the presence of these connections. We revised 
the linkages that were supplied in the previous research by including the five additional $\gamma$\,Dor 
variables that were discovered in our investigation. The updated relations reveal that the connections 
between the pulsation period and the orbital period, the fractional radius of the pulsating component, and 
the force per gram that is exerted on the surface of the pulsating star is less, but they are still 
significant. Detailed spectroscopic investigations of these systems are needed to acquire exact stellar 
parameters, particularly for the $\gamma$\,Dor primary components. This would allow a correct analysis of 
pulsation period relations and modelling of $\gamma$\,Dor's inner structure by asteroseismology, which 
relies on exact stellar parameters. 

\section{acknowledgements}
We thank the anonymous reviewer for their careful reading of our manuscript and their many insightful 
comments and suggestions. This study was funded by Scientific Research Projects Coordination Unit of Istanbul 
University. Project number: FBA-2024-40612. This research made use of data collected at SDSS. Funding 
for the Sloan Digital Sky Survey IV has been provided by the Alfred P. Sloan Foundation, the U.S. 
Department of Energy Office of Science, and the Participating Institutions. SDSS acknowledges 
support and resources from the Center for High-Performance Computing at the University of Utah. The 
SDSS web site is \url{http://www.sdss.org/}. The following internet-based resources were used in 
research for this paper: the NASA Astrophysics Data System; the SIMBAD database operated at CDS, as 
well as of the open-source Python packages ASTROPY {\url{http://www.astropy.org}} \citep{2013A&A...558A..33A}, 
HEALPY {\url{http://healpix.sf.net}} \citep{2005ApJ...622..759G}, MATPLOTLIB \citep{2007CSE.....9...90H}, 
NUMPY \citep{2020Natur.585..357H}, and PANDAS \citep{2020zndo...3630805R}. This work is based on data 
from the \kep\ mission. \kep\ was competitively selected as the tenth Discovery mission. Funding for 
this mission is provided by NASA's Science Mission Directorate. The photometric data were obtained 
from the Mikulski Archive for Space Telescopes (MAST). This work has also made use of data from the 
European Space Agency (ESA) mission \gaia\ (\url{http://www.cosmos.esa.int/gaia}), processed by the 
\gaia\ Data Processing and Analysis Consortium (DPAC, \url{http://www.cosmos.esa.int/web/gaia/dpac/consortium}). This 
research has also made use of "Aladin sky atlas" developed at CDS, Strasbourg Observatory, France.

\section*{Data Availability}
Photometric and spectroscopic raw data used in this paper are publicly available at the
\url{https://archive.stsci.edu/missions-and-data/tess}, and \url{http://archive.eso.org/cms.html} archives.

\bibliographystyle{mnras}
\bibliography{Cakirli_et_al}

\begin{thebibliography}{}
\makeatletter
\relax
\def\mn@urlcharsother{\let\do\@makeother \do\$\do\&\do\#\do\^\do\_\do\%\do\~}
\def\mn@doi{\begingroup\mn@urlcharsother \@ifnextchar [ {\mn@doi@} {\mn@doi@[]}}
\def\mn@doi@[#1]#2{\def\@tempa{#1}\ifx\@tempa\@empty \href {http://dx.doi.org/#2} {doi:#2}\else \href {http://dx.doi.org/#2} {#1}\fi \endgroup}
\def\mn@eprint#1#2{\mn@eprint@#1:#2::\@nil}
\def\mn@eprint@arXiv#1{\href {http://arxiv.org/abs/#1} {{\tt arXiv:#1}}}
\def\mn@eprint@dblp#1{\href {http://dblp.uni-trier.de/rec/bibtex/#1.xml} {dblp:#1}}
\def\mn@eprint@#1:#2:#3:#4\@nil{\def\@tempa {#1}\def\@tempb {#2}\def\@tempc {#3}\ifx \@tempc \@empty \let \@tempc \@tempb \let \@tempb \@tempa \fi \ifx \@tempb \@empty \def\@tempb {arXiv}\fi \@ifundefined {mn@eprint@\@tempb}{\@tempb:\@tempc}{\expandafter \expandafter \csname mn@eprint@\@tempb\endcsname \expandafter{\@tempc}}}

\bibitem[\protect\citeauthoryear{{Aerts}}{{Aerts}}{2021}]{2021RvMP...93a5001A}
{Aerts} C.,  2021, \mn@doi [Reviews of Modern Physics] {10.1103/RevModPhys.93.015001}, \href {https://ui.adsabs.harvard.edu/abs/2021RvMP...93a5001A} {93, 015001}

\bibitem[\protect\citeauthoryear{{Aerts}, {Christensen-Dalsgaard}  \& {Kurtz}}{{Aerts} et~al.}{2010}]{2010aste.book.....A}
{Aerts} C.,  {Christensen-Dalsgaard} J.,   {Kurtz} D.~W.,  2010, {Asteroseismology}, \mn@doi{10.1007/978-1-4020-5803-5.
}

\bibitem[\protect\citeauthoryear{{Astropy Collaboration} et~al.,}{{Astropy Collaboration} et~al.}{2013}]{2013A&A...558A..33A}
{Astropy Collaboration} et~al., 2013, \mn@doi [\aap] {10.1051/0004-6361/201322068}, \href {https://ui.adsabs.harvard.edu/abs/2013A&A...558A..33A} {558, A33}

\bibitem[\protect\citeauthoryear{{Balona}, {Krisciunas}  \& {Cousins}}{{Balona} et~al.}{1994}]{1994MNRAS.270..905B}
{Balona} L.~A.,  {Krisciunas} K.,   {Cousins} A.~W.~J.,  1994, \mn@doi [\mnras] {10.1093/mnras/270.4.905}, \href {https://ui.adsabs.harvard.edu/abs/1994MNRAS.270..905B} {270, 905}

\bibitem[\protect\citeauthoryear{{Balona} et~al.,}{{Balona} et~al.}{2016}]{2016MNRAS.460.1318B}
{Balona} L.~A.,  et~al., 2016, \mn@doi [\mnras] {10.1093/mnras/stw1038}, \href {https://ui.adsabs.harvard.edu/abs/2016MNRAS.460.1318B} {460, 1318}

\bibitem[\protect\citeauthoryear{{Blanco-Cuaresma}}{{Blanco-Cuaresma}}{2019}]{2019MNRAS.486.2075B}
{Blanco-Cuaresma} S.,  2019, \mn@doi [\mnras] {10.1093/mnras/stz549}, \href {https://ui.adsabs.harvard.edu/abs/2019MNRAS.486.2075B} {486, 2075}

\bibitem[\protect\citeauthoryear{{Blanco-Cuaresma}, {Soubiran}, {Heiter}  \& {Jofr{\'e}}}{{Blanco-Cuaresma} et~al.}{2014}]{2014A&A...569A.111B}
{Blanco-Cuaresma} S.,  {Soubiran} C.,  {Heiter} U.,   {Jofr{\'e}} P.,  2014, \mn@doi [\aap] {10.1051/0004-6361/201423945}, \href {https://ui.adsabs.harvard.edu/abs/2014A&A...569A.111B} {569, A111}

\bibitem[\protect\citeauthoryear{{Borucki} et~al.,}{{Borucki} et~al.}{2010}]{2010Sci...327..977B}
{Borucki} W.~J.,  et~al., 2010, \mn@doi [Science] {10.1126/science.1185402}, \href {https://ui.adsabs.harvard.edu/abs/2010Sci...327..977B} {327, 977}

\bibitem[\protect\citeauthoryear{{Bouabid}, {Montalb{\'a}n}, {Miglio}, {Dupret}, {Grigahc{\`e}ne}  \& {Noels}}{{Bouabid} et~al.}{2009}]{2009AIPC.1170..477B}
{Bouabid} M.~P.,  {Montalb{\'a}n} J.,  {Miglio} A.,  {Dupret} M.~A.,  {Grigahc{\`e}ne} A.,   {Noels} A.,  2009, in {Guzik} J.~A.,  {Bradley} P.~A.,  eds,  American Institute of Physics Conference Series Vol. 1170, Stellar Pulsation: Challenges for Theory and Observation. pp 477--479 (\mn@eprint {arXiv} {0911.0775}), \mn@doi{10.1063/1.3246547}

\bibitem[\protect\citeauthoryear{{Bouabid}, {Dupret}, {Salmon}, {Montalb{\'a}n}, {Miglio}  \& {Noels}}{{Bouabid} et~al.}{2013}]{2013MNRAS.429.2500B}
{Bouabid} M.~P.,  {Dupret} M.~A.,  {Salmon} S.,  {Montalb{\'a}n} J.,  {Miglio} A.,   {Noels} A.,  2013, \mn@doi [\mnras] {10.1093/mnras/sts517}, \href {https://ui.adsabs.harvard.edu/abs/2013MNRAS.429.2500B} {429, 2500}

\bibitem[\protect\citeauthoryear{{Castelli} \& {Kurucz}}{{Castelli} \& {Kurucz}}{2003}]{2003IAUS..210P.A20C}
{Castelli} F.,  {Kurucz} R.~L.,  2003, in {Piskunov} N.,  {Weiss} W.~W.,   {Gray} D.~F.,  eds,  Proceedings of the 210th Symposium of the International Astronomical Union Vol. 210, Modelling of Stellar Atmospheres. p.~A20 (\mn@eprint {arXiv} {astro-ph/0405087})

\bibitem[\protect\citeauthoryear{{Cheng}, {Fuller}, {Guo}, {Lehman}  \& {Hambleton}}{{Cheng} et~al.}{2020}]{2020ApJ...903..122C}
{Cheng} S.~J.,  {Fuller} J.,  {Guo} Z.,  {Lehman} H.,   {Hambleton} K.,  2020, \mn@doi [\apj] {10.3847/1538-4357/abb46d}, \href {https://ui.adsabs.harvard.edu/abs/2020ApJ...903..122C} {903, 122}

\bibitem[\protect\citeauthoryear{{Choi}, {Dotter}, {Conroy}, {Cantiello}, {Paxton}  \& {Johnson}}{{Choi} et~al.}{2016}]{2016ApJ...823..102C}
{Choi} J.,  {Dotter} A.,  {Conroy} C.,  {Cantiello} M.,  {Paxton} B.,   {Johnson} B.~D.,  2016, \mn@doi [\apj] {10.3847/0004-637X/823/2/102}, \href {https://ui.adsabs.harvard.edu/abs/2016ApJ...823..102C} {823, 102}

\bibitem[\protect\citeauthoryear{{Clark Cunningham}, {Rawls}, {Windemuth}, {Ali}, {Jackiewicz}, {Agol}  \& {Stassun}}{{Clark Cunningham} et~al.}{2019}]{2019AJ....158..106C}
{Clark Cunningham} J.~M.,  {Rawls} M.~L.,  {Windemuth} D.,  {Ali} A.,  {Jackiewicz} J.,  {Agol} E.,   {Stassun} K.~G.,  2019, \mn@doi [\aj] {10.3847/1538-3881/ab2d2b}, \href {https://ui.adsabs.harvard.edu/abs/2019AJ....158..106C} {158, 106}

\bibitem[\protect\citeauthoryear{{Cox} \& {Pilachowski}}{{Cox} \& {Pilachowski}}{2000}]{2000PhT....53j..77C}
{Cox} A.~N.,  {Pilachowski} C.~A.,  2000, \mn@doi [Physics Today] {10.1063/1.1325201}, \href {https://ui.adsabs.harvard.edu/abs/2000PhT....53j..77C} {53, 77}

\bibitem[\protect\citeauthoryear{{Dekker}, {D'Odorico}, {Kaufer}, {Delabre}  \& {Kotzlowski}}{{Dekker} et~al.}{2000}]{2000SPIE.4008..534D}
{Dekker} H.,  {D'Odorico} S.,  {Kaufer} A.,  {Delabre} B.,   {Kotzlowski} H.,  2000, in {Iye} M.,  {Moorwood} A.~F.,  eds,  Society of Photo-Optical Instrumentation Engineers (SPIE) Conference Series Vol. 4008, Optical and IR Telescope Instrumentation and Detectors. pp 534--545, \mn@doi{10.1117/12.395512}

\bibitem[\protect\citeauthoryear{{Dotter}}{{Dotter}}{2016}]{2016ApJS..222....8D}
{Dotter} A.,  2016, \mn@doi [\apjs] {10.3847/0067-0049/222/1/8}, \href {https://ui.adsabs.harvard.edu/abs/2016ApJS..222....8D} {222, 8}

\bibitem[\protect\citeauthoryear{{Dupret}, {Grigahc{\`e}ne}, {Garrido}, {Gabriel}  \& {Scuflaire}}{{Dupret} et~al.}{2005}]{2005A&A...435..927D}
{Dupret} M.~A.,  {Grigahc{\`e}ne} A.,  {Garrido} R.,  {Gabriel} M.,   {Scuflaire} R.,  2005, \mn@doi [\aap] {10.1051/0004-6361:20041817}, \href {https://ui.adsabs.harvard.edu/abs/2005A&A...435..927D} {435, 927}

\bibitem[\protect\citeauthoryear{{Fernie}}{{Fernie}}{1995}]{1995AJ....110.2361F}
{Fernie} J.~D.,  1995, \mn@doi [\aj] {10.1086/117693}, \href {https://ui.adsabs.harvard.edu/abs/1995AJ....110.2361F} {110, 2361}

\bibitem[\protect\citeauthoryear{{Fuller} \& {Lai}}{{Fuller} \& {Lai}}{2012}]{2012MNRAS.420.3126F}
{Fuller} J.,  {Lai} D.,  2012, \mn@doi [\mnras] {10.1111/j.1365-2966.2011.20237.x}, \href {https://ui.adsabs.harvard.edu/abs/2012MNRAS.420.3126F} {420, 3126}

\bibitem[\protect\citeauthoryear{{Gaia Collaboration} et~al.,}{{Gaia Collaboration} et~al.}{2023}]{2023A&A...674A...1G}
{Gaia Collaboration} et~al., 2023, \mn@doi [\aap] {10.1051/0004-6361/202243940}, \href {https://ui.adsabs.harvard.edu/abs/2023A&A...674A...1G} {674, A1}

\bibitem[\protect\citeauthoryear{{Gaulme} \& {Guzik}}{{Gaulme} \& {Guzik}}{2019}]{2019A&A...630A.106G}
{Gaulme} P.,  {Guzik} J.~A.,  2019, \mn@doi [\aap] {10.1051/0004-6361/201935821}, \href {https://ui.adsabs.harvard.edu/abs/2019A&A...630A.106G} {630, A106}

\bibitem[\protect\citeauthoryear{{Goodman} \& {Weare}}{{Goodman} \& {Weare}}{2010}]{2010CAMCS...5...65G}
{Goodman} J.,  {Weare} J.,  2010, \mn@doi [Communications in Applied Mathematics and Computational Science] {10.2140/camcos.2010.5.65}, \href {https://ui.adsabs.harvard.edu/abs/2010CAMCS...5...65G} {5, 65}

\bibitem[\protect\citeauthoryear{{G{\'o}rski}, {Hivon}, {Banday}, {Wandelt}, {Hansen}, {Reinecke}  \& {Bartelmann}}{{G{\'o}rski} et~al.}{2005}]{2005ApJ...622..759G}
{G{\'o}rski} K.~M.,  {Hivon} E.,  {Banday} A.~J.,  {Wandelt} B.~D.,  {Hansen} F.~K.,  {Reinecke} M.,   {Bartelmann} M.,  2005, \mn@doi [\apj] {10.1086/427976}, \href {https://ui.adsabs.harvard.edu/abs/2005ApJ...622..759G} {622, 759}

\bibitem[\protect\citeauthoryear{{Gray} \& {Corbally}}{{Gray} \& {Corbally}}{1994}]{1994AJ....107..742G}
{Gray} R.~O.,  {Corbally} C.~J.,  1994, \mn@doi [\aj] {10.1086/116893}, \href {https://ui.adsabs.harvard.edu/abs/1994AJ....107..742G} {107, 742}

\bibitem[\protect\citeauthoryear{{Gray} \& {Nagel}}{{Gray} \& {Nagel}}{1989}]{1989ApJ...341..421G}
{Gray} D.~F.,  {Nagel} T.,  1989, \mn@doi [\apj] {10.1086/167505}, \href {https://ui.adsabs.harvard.edu/abs/1989ApJ...341..421G} {341, 421}

\bibitem[\protect\citeauthoryear{{Gunn} et~al.,}{{Gunn} et~al.}{2006}]{2006AJ....131.2332G}
{Gunn} J.~E.,  et~al., 2006, \mn@doi [\aj] {10.1086/500975}, \href {https://ui.adsabs.harvard.edu/abs/2006AJ....131.2332G} {131, 2332}

\bibitem[\protect\citeauthoryear{{G{\"u}zel} \& {{\"O}zdarcan}}{{G{\"u}zel} \& {{\"O}zdarcan}}{2020}]{PyWD2015_2020CoSka..50..535G}
{G{\"u}zel} O.,  {{\"O}zdarcan} O.,  2020, \mn@doi [Contributions of the Astronomical Observatory Skalnate Pleso] {10.31577/caosp.2020.50.2.535}, \href {https://ui.adsabs.harvard.edu/abs/2020CoSka..50..535G} {50, 535}

\bibitem[\protect\citeauthoryear{{Guzik}, {Kaye}, {Bradley}, {Cox}  \& {Neuforge}}{{Guzik} et~al.}{2000}]{2000ApJ...542L..57G}
{Guzik} J.~A.,  {Kaye} A.~B.,  {Bradley} P.~A.,  {Cox} A.~N.,   {Neuforge} C.,  2000, \mn@doi [\apjl] {10.1086/312908}, \href {https://ui.adsabs.harvard.edu/abs/2000ApJ...542L..57G} {542, L57}

\bibitem[\protect\citeauthoryear{{Handler}}{{Handler}}{2013}]{2013pss4.book..207H}
{Handler} G.,  2013, in {Oswalt} T.~D.,  {Barstow} M.~A.,  eds, , Vol.~4, Planets, Stars and Stellar Systems. Volume 4: Stellar Structure and Evolution.
Springer Dordrecht, p.~207, \mn@doi{10.1007/978-94-007-5615-1\_4}

\bibitem[\protect\citeauthoryear{{Handler} \& {Shobbrook}}{{Handler} \& {Shobbrook}}{2002}]{2002MNRAS.333..251H}
{Handler} G.,  {Shobbrook} R.~R.,  2002, \mn@doi [\mnras] {10.1046/j.1365-8711.2002.05401.x}, \href {https://ui.adsabs.harvard.edu/abs/2002MNRAS.333..251H} {333, 251}

\bibitem[\protect\citeauthoryear{{Harris} et~al.,}{{Harris} et~al.}{2020}]{2020Natur.585..357H}
{Harris} C.~R.,  et~al., 2020, \mn@doi [\nat] {10.1038/s41586-020-2649-2}, \href {https://ui.adsabs.harvard.edu/abs/2020Natur.585..357H} {585, 357}

\bibitem[\protect\citeauthoryear{{Henry}, {Fekel}  \& {Williamson}}{{Henry} et~al.}{2022}]{2022AJ....163..180H}
{Henry} G.~W.,  {Fekel} F.~C.,   {Williamson} M.~H.,  2022, \mn@doi [\aj] {10.3847/1538-3881/ac540b}, \href {https://ui.adsabs.harvard.edu/abs/2022AJ....163..180H} {163, 180}

\bibitem[\protect\citeauthoryear{{Hogg} \& {Foreman-Mackey}}{{Hogg} \& {Foreman-Mackey}}{2018}]{2018ApJS..236...11H}
{Hogg} D.~W.,  {Foreman-Mackey} D.,  2018, \mn@doi [\apjs] {10.3847/1538-4365/aab76e}, \href {https://ui.adsabs.harvard.edu/abs/2018ApJS..236...11H} {236, 11}

\bibitem[\protect\citeauthoryear{{Houdek}, {Balmforth}, {Christensen-Dalsgaard}  \& {Gough}}{{Houdek} et~al.}{1999}]{1999A&A...351..582H}
{Houdek} G.,  {Balmforth} N.~J.,  {Christensen-Dalsgaard} J.,   {Gough} D.~O.,  1999, \aap, \href {https://ui.adsabs.harvard.edu/abs/1999A&A...351..582H} {351, 582}

\bibitem[\protect\citeauthoryear{{Howell} et~al.,}{{Howell} et~al.}{2014}]{2014PASP..126..398H}
{Howell} S.~B.,  et~al., 2014, \mn@doi [\pasp] {10.1086/676406}, \href {https://ui.adsabs.harvard.edu/abs/2014PASP..126..398H} {126, 398}

\bibitem[\protect\citeauthoryear{{Hoyman}, {{\c{C}}ak{\i}rl{\i}}  \& {{\"O}zdarcan}}{{Hoyman} et~al.}{2020}]{2020MNRAS.491.5980H}
{Hoyman} B.,  {{\c{C}}ak{\i}rl{\i}} {\"O}.,   {{\"O}zdarcan} O.,  2020, \mn@doi [\mnras] {10.1093/mnras/stz3302}, \href {https://ui.adsabs.harvard.edu/abs/2020MNRAS.491.5980H} {491, 5980}

\bibitem[\protect\citeauthoryear{{Hunter}}{{Hunter}}{2007}]{2007CSE.....9...90H}
{Hunter} J.~D.,  2007, \mn@doi [Computing in Science and Engineering] {10.1109/MCSE.2007.55}, \href {https://ui.adsabs.harvard.edu/abs/2007CSE.....9...90H} {9, 90}

\bibitem[\protect\citeauthoryear{{Hut}}{{Hut}}{1981}]{1981A&A....99..126H}
{Hut} P.,  1981, \aap, \href {https://ui.adsabs.harvard.edu/abs/1981A&A....99..126H} {99, 126}

\bibitem[\protect\citeauthoryear{{Ibanoglu}, {{\c{C}}ak{\i}rl{\i}}  \& {Sipahi}}{{Ibanoglu} et~al.}{2018}]{2018NewA...62...70I}
{Ibanoglu} C.,  {{\c{C}}ak{\i}rl{\i}} {\"O}.,   {Sipahi} E.,  2018, \mn@doi [\na] {10.1016/j.newast.2018.01.004}, \href {https://ui.adsabs.harvard.edu/abs/2018NewA...62...70I} {62, 70}

\bibitem[\protect\citeauthoryear{{Jenkins} et~al.,}{{Jenkins} et~al.}{2016}]{2016SPIE.9913E..3EJ}
{Jenkins} J.~M.,  et~al., 2016, in {Chiozzi} G.,  {Guzman} J.~C.,  eds,  Society of Photo-Optical Instrumentation Engineers (SPIE) Conference Series Vol. 9913, Software and Cyberinfrastructure for Astronomy IV. p. 99133E, \mn@doi{10.1117/12.2233418}

\bibitem[\protect\citeauthoryear{{Kallinger}, {Reegen}  \& {Weiss}}{{Kallinger} et~al.}{2008}]{2008A&A...481..571K}
{Kallinger} T.,  {Reegen} P.,   {Weiss} W.~W.,  2008, \mn@doi [\aap] {10.1051/0004-6361:20077559}, \href {https://ui.adsabs.harvard.edu/abs/2008A&A...481..571K} {481, 571}

\bibitem[\protect\citeauthoryear{{Kaufer}, {Stahl}, {Tubbesing}, {N{\o}rregaard}, {Avila}, {Francois}, {Pasquini}  \& {Pizzella}}{{Kaufer} et~al.}{1999}]{1999Msngr..95....8K}
{Kaufer} A.,  {Stahl} O.,  {Tubbesing} S.,  {N{\o}rregaard} P.,  {Avila} G.,  {Francois} P.,  {Pasquini} L.,   {Pizzella} A.,  1999, The Messenger, \href {https://ui.adsabs.harvard.edu/abs/1999Msngr..95....8K} {95, 8}

\bibitem[\protect\citeauthoryear{{Kaye}, {Handler}, {Krisciunas}, {Poretti}  \& {Zerbi}}{{Kaye} et~al.}{1999}]{1999PASP..111..840K}
{Kaye} A.~B.,  {Handler} G.,  {Krisciunas} K.,  {Poretti} E.,   {Zerbi} F.~M.,  1999, \mn@doi [\pasp] {10.1086/316399}, \href {https://ui.adsabs.harvard.edu/abs/1999PASP..111..840K} {111, 840}

\bibitem[\protect\citeauthoryear{{Kurtz}, {Saio}, {Takata}, {Shibahashi}, {Murphy}  \& {Sekii}}{{Kurtz} et~al.}{2014}]{2014MNRAS.444..102K}
{Kurtz} D.~W.,  {Saio} H.,  {Takata} M.,  {Shibahashi} H.,  {Murphy} S.~J.,   {Sekii} T.,  2014, \mn@doi [\mnras] {10.1093/mnras/stu1329}, \href {https://ui.adsabs.harvard.edu/abs/2014MNRAS.444..102K} {444, 102}

\bibitem[\protect\citeauthoryear{{Li}, {Van Reeth}, {Bedding}, {Murphy}  \& {Antoci}}{{Li} et~al.}{2019}]{2019MNRAS.487..782L}
{Li} G.,  {Van Reeth} T.,  {Bedding} T.~R.,  {Murphy} S.~J.,   {Antoci} V.,  2019, \mn@doi [\mnras] {10.1093/mnras/stz1171}, \href {https://ui.adsabs.harvard.edu/abs/2019MNRAS.487..782L} {487, 782}

\bibitem[\protect\citeauthoryear{{Liakos}, {Niarchos}, {Soydugan}  \& {Zasche}}{{Liakos} et~al.}{2012}]{2012MNRAS.422.1250L}
{Liakos} A.,  {Niarchos} P.,  {Soydugan} E.,   {Zasche} P.,  2012, \mn@doi [\mnras] {10.1111/j.1365-2966.2012.20704.x}, \href {https://ui.adsabs.harvard.edu/abs/2012MNRAS.422.1250L} {422, 1250}

\bibitem[\protect\citeauthoryear{{Lightkurve Collaboration} et~al.,}{{Lightkurve Collaboration} et~al.}{2018}]{2018ascl.soft12013L}
{Lightkurve Collaboration} et~al., 2018, {Lightkurve: Kepler and TESS time series analysis in Python}, Astrophysics Source Code Library (\mn@eprint {ascl} {1812.013})

\bibitem[\protect\citeauthoryear{{Lucy}}{{Lucy}}{1967}]{1967ZA.....65...89L}
{Lucy} L.~B.,  1967, \zap, \href {https://ui.adsabs.harvard.edu/abs/1967ZA.....65...89L} {65, 89}

\bibitem[\protect\citeauthoryear{{Majewski} et~al.,}{{Majewski} et~al.}{2017}]{2017AJ....154...94M}
{Majewski} S.~R.,  et~al., 2017, \mn@doi [\aj] {10.3847/1538-3881/aa784d}, \href {https://ui.adsabs.harvard.edu/abs/2017AJ....154...94M} {154, 94}

\bibitem[\protect\citeauthoryear{{Matson}, {Gies}, {Guo}  \& {Orosz}}{{Matson} et~al.}{2016}]{2016AJ....151..139M}
{Matson} R.~A.,  {Gies} D.~R.,  {Guo} Z.,   {Orosz} J.~A.,  2016, \mn@doi [\aj] {10.3847/0004-6256/151/6/139}, \href {https://ui.adsabs.harvard.edu/abs/2016AJ....151..139M} {151, 139}

\bibitem[\protect\citeauthoryear{{Mayor} et~al.,}{{Mayor} et~al.}{2003}]{2003Msngr.114...20M}
{Mayor} M.,  et~al., 2003, The Messenger, \href {https://ui.adsabs.harvard.edu/abs/2003Msngr.114...20M} {114, 20}

\bibitem[\protect\citeauthoryear{{Munari} \& {Zwitter}}{{Munari} \& {Zwitter}}{1997}]{1997A&A...318..269M}
{Munari} U.,  {Zwitter} T.,  1997, \aap, \href {https://ui.adsabs.harvard.edu/abs/1997A&A...318..269M} {318, 269}

\bibitem[\protect\citeauthoryear{{Nidever} et~al.,}{{Nidever} et~al.}{2015}]{2015AJ....150..173N}
{Nidever} D.~L.,  et~al., 2015, \mn@doi [\aj] {10.1088/0004-6256/150/6/173}, \href {https://ui.adsabs.harvard.edu/abs/2015AJ....150..173N} {150, 173}

\bibitem[\protect\citeauthoryear{{{\"O}zdarcan} \& {{\c{C}}akirli}}{{{\"O}zdarcan} \& {{\c{C}}akirli}}{2020}]{2020RMxAA..56..321O}
{{\"O}zdarcan} O.,  {{\c{C}}akirli} {\"O}.,  2020, \mn@doi [\rmxaa] {10.22201/ia.01851101p.2020.56.02.12}, \href {https://ui.adsabs.harvard.edu/abs/2020RMxAA..56..321O} {56, 321}

\bibitem[\protect\citeauthoryear{{P{\'a}pics}}{{P{\'a}pics}}{2012}]{2012AN....333.1053P}
{P{\'a}pics} P.~I.,  2012, \mn@doi [Astronomische Nachrichten] {10.1002/asna.201211809}, \href {https://ui.adsabs.harvard.edu/abs/2012AN....333.1053P} {333, 1053}

\bibitem[\protect\citeauthoryear{{Paxton} et~al.,}{{Paxton} et~al.}{2015}]{2015ApJS..220...15P}
{Paxton} B.,  et~al., 2015, \mn@doi [\apjs] {10.1088/0067-0049/220/1/15}, \href {https://ui.adsabs.harvard.edu/abs/2015ApJS..220...15P} {220, 15}

\bibitem[\protect\citeauthoryear{{Pr{\v{s}}a} et~al.,}{{Pr{\v{s}}a} et~al.}{2022}]{2022ApJS..258...16P}
{Pr{\v{s}}a} A.,  et~al., 2022, \mn@doi [\apjs] {10.3847/1538-4365/ac324a}, \href {https://ui.adsabs.harvard.edu/abs/2022ApJS..258...16P} {258, 16}

\bibitem[\protect\citeauthoryear{{Reback} et~al.,}{{Reback} et~al.}{2020}]{2020zndo...3630805R}
{Reback} J.,  et~al., 2020, {pandas-dev/pandas: Pandas 1.0.0}, Zenodo, \mn@doi{10.5281/zenodo.3630805}

\bibitem[\protect\citeauthoryear{{Reegen}}{{Reegen}}{2007}]{2007A&A...467.1353R}
{Reegen} P.,  2007, \mn@doi [\aap] {10.1051/0004-6361:20066597}, \href {https://ui.adsabs.harvard.edu/abs/2007A&A...467.1353R} {467, 1353}

\bibitem[\protect\citeauthoryear{{Ricker} et~al.,}{{Ricker} et~al.}{2015}]{2015JATIS...1a4003R}
{Ricker} G.~R.,  et~al., 2015, \mn@doi [Journal of Astronomical Telescopes, Instruments, and Systems] {10.1117/1.JATIS.1.1.014003}, \href {https://ui.adsabs.harvard.edu/abs/2015JATIS...1a4003R} {1, 014003}

\bibitem[\protect\citeauthoryear{{Ruci{\'n}ski}}{{Ruci{\'n}ski}}{1969}]{1969AcA....19..245R}
{Ruci{\'n}ski} S.~M.,  1969, \actaa, \href {https://ui.adsabs.harvard.edu/abs/1969AcA....19..245R} {19, 245}

\bibitem[\protect\citeauthoryear{{Saio}, {Kurtz}, {Murphy}, {Antoci}  \& {Lee}}{{Saio} et~al.}{2018a}]{2018MNRAS.474.2774S}
{Saio} H.,  {Kurtz} D.~W.,  {Murphy} S.~J.,  {Antoci} V.~L.,   {Lee} U.,  2018a, \mn@doi [\mnras] {10.1093/mnras/stx2962}, \href {https://ui.adsabs.harvard.edu/abs/2018MNRAS.474.2774S} {474, 2774}

\bibitem[\protect\citeauthoryear{{Saio}, {Bedding}, {Kurtz}, {Murphy}, {Antoci}, {Shibahashi}, {Li}  \& {Takata}}{{Saio} et~al.}{2018b}]{2018MNRAS.477.2183S}
{Saio} H.,  {Bedding} T.~R.,  {Kurtz} D.~W.,  {Murphy} S.~J.,  {Antoci} V.,  {Shibahashi} H.,  {Li} G.,   {Takata} M.,  2018b, \mn@doi [\mnras] {10.1093/mnras/sty784}, \href {https://ui.adsabs.harvard.edu/abs/2018MNRAS.477.2183S} {477, 2183}

\bibitem[\protect\citeauthoryear{{Samadi}, {Goupil}  \& {Houdek}}{{Samadi} et~al.}{2002}]{2002A&A...395..563S}
{Samadi} R.,  {Goupil} M.~J.,   {Houdek} G.,  2002, \mn@doi [\aap] {10.1051/0004-6361:20021322}, \href {https://ui.adsabs.harvard.edu/abs/2002A&A...395..563S} {395, 563}

\bibitem[\protect\citeauthoryear{{Sekaran} et~al.,}{{Sekaran} et~al.}{2020}]{2020A&A...643A.162S}
{Sekaran} S.,  et~al., 2020, \mn@doi [\aap] {10.1051/0004-6361/202038989}, \href {https://ui.adsabs.harvard.edu/abs/2020A&A...643A.162S} {643, A162}

\bibitem[\protect\citeauthoryear{{Shporer} et~al.,}{{Shporer} et~al.}{2016}]{2016ApJ...829...34S}
{Shporer} A.,  et~al., 2016, \mn@doi [\apj] {10.3847/0004-637X/829/1/34}, \href {https://ui.adsabs.harvard.edu/abs/2016ApJ...829...34S} {829, 34}

\bibitem[\protect\citeauthoryear{{Southworth}, {Maxted}  \& {Smalley}}{{Southworth} et~al.}{2005}]{2005A&A...429..645S}
{Southworth} J.,  {Maxted} P.~F.~L.,   {Smalley} B.,  2005, \mn@doi [\aap] {10.1051/0004-6361:20041867}, \href {https://ui.adsabs.harvard.edu/abs/2005A&A...429..645S} {429, 645}

\bibitem[\protect\citeauthoryear{{Soydugan}, {{\.I}bano{\v{g}}lu}, {Soydugan}, {Akan}  \& {Demircan}}{{Soydugan} et~al.}{2006}]{2006MNRAS.366.1289S}
{Soydugan} E.,  {{\.I}bano{\v{g}}lu} C.,  {Soydugan} F.,  {Akan} M.~C.,   {Demircan} O.,  2006, \mn@doi [\mnras] {10.1111/j.1365-2966.2005.09889.x}, \href {https://ui.adsabs.harvard.edu/abs/2006MNRAS.366.1289S} {366, 1289}

\bibitem[\protect\citeauthoryear{{Tout}, {Pols}, {Eggleton}  \& {Han}}{{Tout} et~al.}{1996}]{1996MNRAS.281..257T}
{Tout} C.~A.,  {Pols} O.~R.,  {Eggleton} P.~P.,   {Han} Z.,  1996, \mn@doi [\mnras] {10.1093/mnras/281.1.257}, \href {https://ui.adsabs.harvard.edu/abs/1996MNRAS.281..257T} {281, 257}

\bibitem[\protect\citeauthoryear{{Triana}, {Moravveji}, {P{\'a}pics}, {Aerts}, {Kawaler}  \& {Christensen-Dalsgaard}}{{Triana} et~al.}{2015}]{2015ApJ...810...16T}
{Triana} S.~A.,  {Moravveji} E.,  {P{\'a}pics} P.~I.,  {Aerts} C.,  {Kawaler} S.~D.,   {Christensen-Dalsgaard} J.,  2015, \mn@doi [\apj] {10.1088/0004-637X/810/1/16}, \href {https://ui.adsabs.harvard.edu/abs/2015ApJ...810...16T} {810, 16}

\bibitem[\protect\citeauthoryear{{Van Reeth}, {Tkachenko}  \& {Aerts}}{{Van Reeth} et~al.}{2016}]{2016A&A...593A.120V}
{Van Reeth} T.,  {Tkachenko} A.,   {Aerts} C.,  2016, \mn@doi [\aap] {10.1051/0004-6361/201628616}, \href {https://ui.adsabs.harvard.edu/abs/2016A&A...593A.120V} {593, A120}

\bibitem[\protect\citeauthoryear{{Van Reeth} et~al.,}{{Van Reeth} et~al.}{2018}]{2018A&A...618A..24V}
{Van Reeth} T.,  et~al., 2018, \mn@doi [\aap] {10.1051/0004-6361/201832718}, \href {https://ui.adsabs.harvard.edu/abs/2018A&A...618A..24V} {618, A24}

\bibitem[\protect\citeauthoryear{{Warner}, {Kaye}  \& {Guzik}}{{Warner} et~al.}{2003}]{2003ApJ...593.1049W}
{Warner} P.~B.,  {Kaye} A.~B.,   {Guzik} J.~A.,  2003, \mn@doi [\apj] {10.1086/376727}, \href {https://ui.adsabs.harvard.edu/abs/2003ApJ...593.1049W} {593, 1049}

\bibitem[\protect\citeauthoryear{{Wilson} \& {Devinney}}{{Wilson} \& {Devinney}}{1971}]{WD_MAIN_1971ApJ}
{Wilson} R.~E.,  {Devinney} E.~J.,  1971, \mn@doi [\apj] {10.1086/150986}, \href {http://adsabs.harvard.edu/abs/1971ApJ...166..605W} {166, 605}

\bibitem[\protect\citeauthoryear{{Wilson} \& {Van Hamme}}{{Wilson} \& {Van Hamme}}{2014}]{WD2015_2014ApJ}
{Wilson} R.~E.,  {Van Hamme} W.,  2014, \mn@doi [\apj] {10.1088/0004-637X/780/2/151}, \href {http://adsabs.harvard.edu/abs/2014ApJ...780..151W} {780, 151}

\bibitem[\protect\citeauthoryear{{Xiong}, {Deng}, {Zhang}  \& {Wang}}{{Xiong} et~al.}{2016}]{2016MNRAS.457.3163X}
{Xiong} D.~R.,  {Deng} L.,  {Zhang} C.,   {Wang} K.,  2016, \mn@doi [\mnras] {10.1093/mnras/stw047}, \href {https://ui.adsabs.harvard.edu/abs/2016MNRAS.457.3163X} {457, 3163}

\bibitem[\protect\citeauthoryear{{{\c{C}}ak{\i}rl{\i}} \& {Ibanoglu}}{{{\c{C}}ak{\i}rl{\i}} \& {Ibanoglu}}{2016}]{2016NewA...45...36C}
{{\c{C}}ak{\i}rl{\i}} {\"O}.,  {Ibanoglu} C.,  2016, \mn@doi [\na] {10.1016/j.newast.2015.10.003}, \href {https://ui.adsabs.harvard.edu/abs/2016NewA...45...36C} {45, 36}

\bibitem[\protect\citeauthoryear{{{\c{C}}ak{\i}rl{\i}}, {{\"O}zdarcan}  \& {Hoyman}}{{{\c{C}}ak{\i}rl{\i}} et~al.}{2023}]{2023MNRAS.526.5987C}
{{\c{C}}ak{\i}rl{\i}} {\"O}.,  {{\"O}zdarcan} O.,   {Hoyman} B.,  2023, \mn@doi [\mnras] {10.1093/mnras/stad3146}, \href {https://ui.adsabs.harvard.edu/abs/2023MNRAS.526.5987C} {526, 5987}

\bibitem[\protect\citeauthoryear{{van Hamme}}{{van Hamme}}{1993}]{1993AJ....106.2096V}
{van Hamme} W.,  1993, \mn@doi [\aj] {10.1086/116788}, \href {https://ui.adsabs.harvard.edu/abs/1993AJ....106.2096V} {106, 2096}

\makeatother
\end{thebibliography}


\appendix
\section{Radial velocity measurements of targets}
\fontsize{9}{11}\selectfont
\onecolumn
\begin{longtable}{l c c c c c c c c}
	\caption{Log of the radial velocity measurements of the targets.}
	\label{tab:target_RVs}\\
	\toprule
	\makecell[b]{System\\~}
	& \makecell[b]{HJD\\(+2400000)}
	& \makecell[b]{$v_1$\\(\kms)}
	& \makecell[b]{$\sigma_1$\\(\kms)}
	& \makecell[b]{$v_2$\\(\kms)}
	& \makecell[b]{$\sigma_2$\\(\kms)}
	& \makecell[b]{S/N$^a$\\~}
	& \makecell[b]{Instrument\\~}\\
	
	\midrule
	\endfirsthead
	\caption{(Cont.)}\\
	\toprule
	
	\makecell[b]{System\\~}
	&\makecell[b]{HJD\\(+2400000)}
	& \makecell[b]{$v_1$\\(\kms)}
	& \makecell[b]{$\sigma_1$\\(\kms)}
	& \makecell[b]{$v_2$\\(\kms)}
	& \makecell[b]{$\sigma_2$\\(\kms)}
	& \makecell[b]{S/N$^a$\\~}
	& \makecell[b]{Instrument\\~}\\
	\midrule
	\endhead
	\midrule
	\multicolumn{8}{r}{\footnotesize\itshape Continue on the next page}
	\endfoot
	\bottomrule
	\multicolumn{8}{l}{\footnotesize $^a$ S/N values have been obtained from headers.}
	\endlastfoot
KIC\,4932691&55813.7025 &	128.9 & 4.3 &  49.6  &2.9 & 33 & APOGEE \\
			&55823.7264 &	 64.0 & 4.2 & 112.2  &5.1 & 44 & APOGEE \\
			&55840.6611 &	 26.5 & 7.1 & 142.5  &2.9 & 55 & APOGEE \\
			&55849.5783 &	133.4 & 5.7 &  47.5  &3.7 & 61 & APOGEE \\
			&55851.6487 &	121.9 & 3.5 &  48.9  &3.8 & 33 & APOGEE \\
			&55866.5694 &	127.5 & 5.3 &  52.7  &3.6 & 25 & APOGEE \\
	\midrule
TIC\,7695666&58097.6645  &   -50.1 &     0.2  &   106.1 &	  0.7   & 62 &  HARPS  \\
			&58098.6225  &	  -8.5 &     0.2  &    54.5 &	  0.7	& 44 &  HARPS  \\
			&58099.6207  &	  21.2 &     0.2  &   ---   &	   ---	& 46 &  HARPS  \\
			&58139.5499  &	  50.1 &     0.3  &    -2.0 &	  0.4   & 53 &  HARPS  \\
			&58140.5773  &	  52.0 &     0.2  &    -4.0 &	  0.5	& 58 &  HARPS  \\
			&59167.6196  &	 -69.3 &     0.5  &   127.0 &	  0.1	& 71 &  HARPS  \\
			&59167.7893  &	 -72.8 &     0.4  &   130.6 &	  0.3	& 22 &  HARPS  \\
			&59510.7613  &	  52.8 &     0.7  &    -4.9 &	  0.1	& 22 &  HARPS  \\
			&59511.6908  &	  52.8 &     0.6  &    -4.9 &	  0.3	& 27 &  HARPS  \\
			&59512.7168  &	  51.9 &     0.5  &    -3.8 &	  0.8	& 27 &  HARPS  \\
			&59513.7444  &	  49.4 &     0.5  &    -0.9 &	  0.3	& 35 &  HARPS  \\	
	\midrule
TIC\,410418820&54807.59791 &    8.5 &  0.2 &     -14.9 &  0.4  &150 &FEROS \\
		&54808.60247 &   50.8 &  0.1 &     -81.1 &  0.8  &120 &FEROS \\
		&55145.74362 &	  5.7 &  0.0 &     -18.3 &  1.2  &120 &FEROS \\
		&55192.84014 &   -8.8 &  0.2 &      12.9 &  2.7  &120 &FEROS \\
		&55192.84282 &   -8.8 &  0.3 &      12.4 &  0.7  &120 &FEROS \\
		&55193.56466 &   26.0 &  0.1 &     -42.3 &  1.2  &120 &FEROS \\
		&55193.56620 &   26.2 &  0.1 &     -43.3 &  1.3  &120 &FEROS \\
		&55193.56795 &   26.2 &  0.1 &     -42.7 &  0.2  &100 &FEROS \\
		&55193.84988 &   38.1 &  0.1 &     -60.7 &  0.2  &100 &FEROS \\
		&55194.85220 &   63.7 &  0.1 &    -106.9 &  0.3  &180 &FEROS \\
		&55197.56061 &  -11.4 &  0.1 &      20.9 &  1.3  &180 &FEROS \\
		&55197.70021 &  -16.8 &  0.1 &      32.4 &  1.6  &180 &FEROS \\
		&55201.73883 &	  8.0 &  0.3 &     -12.2 &  0.7  &180 &FEROS \\
		&55477.90044 &   67.3 &  0.2 &    -104.5 &  0.2  &180 &FEROS \\
		&55477.90044 &   64.8 &  0.1 &    -104.5 &  2.1  &180 &HARPS \\
		&55479.73996 &   15.3 &  0.2 &     -21.1 &  2.1  &180 &HARPS \\
		&55479.73996 &   13.7 &  0.1 &     -20.3 &  1.6  &40  &HARPS \\
		&56214.87621 &   67.1 &  0.2 &    -103.9 &  1.2  &120 &HARPS \\
		&56214.87621 &   65.1 &  0.1 &    -103.9 &  1.2  &80  &HARPS \\
		&56579.90006 &  -47.3 &  0.2 &      86.7 &  2.1  &180 &HARPS \\
		&56579.90006 &  -50.6 &  0.1 &      85.8 &  1.7  &180 &HARPS \\
		&56908.71869 &   65.0 &  0.1 &    -102.1 &  1.7  &180 &HARPS \\
		&56909.86544 &   48.2 &  0.1 &     -77.0 &  1.9  &180 &HARPS \\
	\midrule
EPIC\,203929178&57182.67049 &   -128.5 &  0.2 &    96.4  &   0.2 &25 &FEROS \\
				&57183.53460 &	   64.4 &  0.4 &  -110.0  &   0.8 &20 &FEROS \\
				&57184.58284 &	  -80.0 &  0.5 &    44.2  &   0.8 &27 &FEROS \\
				&57185.66882 &	   15.2 &  0.3 &   -65.7  &   0.8 &25 &FEROS \\
				&57186.67824 &	  -20.3 &  0.7 &    ---   &   --- &19 &FEROS \\
				&57187.61361 &	  -79.9 &  0.7 &    44.3  &   0.4 &29 &FEROS \\
				&57191.54023 &	  -87.6 &  0.6 &    52.0  &   0.8 &22 &FEROS \\
				&57193.74259 &	  -52.0 &  0.9 &    26.8  &   0.9 &30 &FEROS \\
				&57194.54276 &	  -77.7 &  0.6 &    41.8  &   0.8 &28 &FEROS \\
				&57195.58909 &	   62.8 &  0.6 &  -108.8  &   0.8 &16 &FEROS \\
				&57211.66564 &	   75.8 &  0.7 &  -123.8  &   0.8 &26 &FEROS \\
				&57584.58852 &	  -68.9 &  0.9 &    38.1  &   0.9 &32 &FEROS \\
				&57584.72573 &	  -46.1 &  0.5 &     4.8  &   1.3 &23 &FEROS \\
	\midrule
TIC\,213047427&57344.73726 &   83.5&   0.7 &   -59.4 &  1.2  &94  &UVES\\
			&57344.74259 &   83.0&   0.5 &   -59.7 &  2.2  &27  &UVES\\
			&57347.70333 &   -5.3&   0.5 &    53.4 &  1.3  &54  &UVES\\
			&57348.71300 &  -34.1&   0.5 &    93.2 &  1.4  &110 &UVES\\
			&57377.65890 &   12.1&   0.5 &    ---  & ---   &103 &UVES\\
			&57377.66574 &   12.2&   0.2 &    ---  & ---   &85  &UVES\\
			&57439.58903 &   26.7&   0.4 &    ---  & ---   &16  &UVES\\
			&57441.52817 &   81.8&   0.2 &   -59.9 &  1.2  &106 &UVES\\
			&57441.53579 &   81.8&   0.3 &   -59.8 &  0.8  &125 &UVES\\
			&57442.54492 &   67.0&   0.4 &   -45.2 &  0.9  &129 &UVES\\
			&57442.55060 &   66.9&   0.2 &   -45.2 &  0.9  &134 &UVES\\
			&57470.50142 &    7.5&   0.4 &    33.7 &  1.2  &157 &UVES
\label{tab:RVs}
\end{longtable}


	\section{Frequencies}
\fontsize{7}{11}\selectfont
\onecolumn
\begin{longtable}{c c c c c}
\caption{Genuine frequencies found in the light curve residuals of the targets.}
\label{tab:Freqs}\\
\toprule
\makecell[b]{ID\\~}
& \makecell[b]{Frequency\\($c/d$)}
& \makecell[b]{$Amplitude$\\($\times10^{-3}$)}
& \makecell[b]{Phase\\(rad)}
& \makecell[b]{RMS\\~}\\

\midrule
\endfirsthead
\caption{(Cont.)}\\
\toprule

\makecell[b]{ID\\~}
& \makecell[b]{Frequency\\($c/d$)}
& \makecell[b]{$Amplitude$\\($\times10^{-3}$)}
& \makecell[b]{Phase\\(rad)}
& \makecell[b]{RMS\\~}\\
\midrule
\endhead
\midrule
\multicolumn{5}{r}{\footnotesize\itshape Continue on the next page} 
\endfoot
\bottomrule
\endlastfoot

 \multicolumn{5}{l}{KIC 4932691} \\ 
\midrule 
$f_{1}$ & 0.983955(13) & 7.6655(29) & 4.7213(90) & 0.0111  \\ 
$f_{2}$ & 1.544509(18) & 5.1581(36) & 0.9042(123) & 0.0097  \\ 
$f_{3}$ & 1.356862(19) & 4.3960(33) & 3.9513(128) & 0.0091  \\ 
$f_{4}$ & 1.479508(18) & 4.4218(30) & 5.9697(122) & 0.0085  \\ 
$f_{5}$ & 1.101472(15) & 4.8037(23) & 5.9890(102) & 0.0079  \\ 
$f_{6}$ & 1.168570(13) & 4.9663(18) & 3.3562(89) & 0.0072  \\ 
$f_{7}$ & 1.698252(15) & 3.7955(18) & 4.7703(100) & 0.0062  \\ 
\midrule 
\multicolumn{5}{l}{TIC 7695666} \\ 
\midrule 
$f_{1}$ & 0.471531(1652) & 0.3054(23) & 5.7696(406) & 0.0011  \\ 
$f_{2}$ & 0.332071(2148) & 0.1554(20) & 3.8411(528) & 0.0011  \\ 
$f_{3}$ & 0.301168(3078) & 0.1414(37) & 3.1630(757) & 0.0011  \\ 
\midrule 
\multicolumn{5}{l}{TIC 410418820} \\ 
\midrule 
$f_{1}$ & 0.679895(1136) & 0.8677(34) & 6.0103(292) & 0.0021  \\ 
$f_{2}$ & 1.762961(1289) & 1.1703(59) & 2.4159(331) & 0.0020  \\ 
$f_{3}$ & 4.842719(1461) & 0.2344(15) & 4.8318(375) & 0.0013  \\ 
$f_{4}$ & 4.348878(1465) & 0.4510(29) & 2.8771(377) & 0.0013  \\ 
\midrule 
\multicolumn{5}{l}{EPIC 203929178} \\ 
\midrule 
$f_{1}$ & 1.121339(592) & 8.4665(178) & 4.2442(214) & 0.0072  \\ 
$f_{2}$ & 1.490798(1061) & 2.6157(177) & 0.7978(383) & 0.0042  \\ 
$f_{3}$ & 0.035293(2402) & 0.9810(340) & 2.0062(867) & 0.0036  \\ 
$f_{4}$ & 0.168287(3997) & 0.5151(494) & 5.0093(1443) & 0.0030  \\ 
$f_{5}$ & 1.653099(4559) & 0.5061(631) & 4.9943(1646) & 0.0028  \\ 
$f_{6}$ & 2.427406(5745) & 0.2570(509) & 2.3411(2074) & 0.0020  \\ 
\midrule 
\multicolumn{5}{l}{TIC 213047427} \\ 
\midrule 
$f_{1}$ & 1.237575(21) & 14.7540(39) & 1.9303(76) & 0.0172  \\ 
$f_{2}$ & 1.437488(20) & 13.6192(30) & 1.5160(70) & 0.0122  \\ 
$f_{3}$ & 2.689875(28) & 9.3779(41) & 1.8750(98) & 0.0078  \\ 
$f_{4}$ & 1.387603(49) & 3.3725(47) & 5.8825(174) & 0.0056  \\ 

\end{longtable}


\bsp	
\label{lastpage}
\end{document}